\documentclass[a4paper,11pt]{article}%
\usepackage{fullpage}
\usepackage[utf8x]{inputenc}
\usepackage{authblk}
\usepackage[breaklinks]{hyperref}
\usepackage[svgnames]{xcolor}
\usepackage{times}
\usepackage{graphicx}
\usepackage{xcolor}
\usepackage{booktabs}
\usepackage{tcolorbox}
\usepackage{caption}
\usepackage[font=small,labelfont=bf]{caption}
\usepackage{amsfonts, amsmath, amsthm, amssymb}
\usepackage{wrapfig}
\usepackage{bbm}
\usepackage{MnSymbol}
\usepackage{multicol}
\usepackage{enumitem}
\usepackage[margin=1.5cm]{geometry}
\usepackage{hyperref}
\usepackage{amsmath}
\usepackage{amsfonts}
\usepackage{caption}
\usepackage{soul} 
\usepackage{amssymb}%
\usepackage{cite}%
\usepackage{ulem}
\usepackage{appendix}
\providecommand{\U}[1]{\protect\rule{.1in}{.1in}}
\numberwithin{equation}{section}
\def\begeq{\begin{equation}}
\def\endeq{\end{equation}}
\newtheorem{lemma}{Lemma}[section]

\newtheorem{theorem}{Theorem}[section]
\newtheorem{proposition}{Proposition}[section]
\newtheorem{definition}{Definition}[section]

\def\begp{\begin{proposition}}
\def\endp{\end{proposition}}
\def\begl{\begin{lemma}}
\def\endl{\end{lemma}}
\def\begt{\begin{theorem}}
\def\endt{\end{theorem}}
\def\begd{\begin{definition}}
\def\endd{\end{definition}}

\newcommand{\eps}{\varepsilon}

\newcommand{\nl}{\newline}

\newcommand{\s}{\mathcal{S}}

\begin{document}

\title{Semi-Device-Independent Random Number Generation with Flexible Assumptions}
\author[1,2]{Matej Pivoluska\footnote{\href{mpivoluska@mail.muni.cz}{mpivoluska@mail.muni.cz}}}
\author[1,2]{Martin Plesch\footnote{\href{plesch@savba.sk}{plesch@savba.sk}}}
\author[3,4,5]{M\'{a}t\'{e} Farkas\footnote{\href{mate.farkas@icfo.eu}{mate.farkas@icfo.eu}}}
\author[6]{Nat\'{a}lia Ru\v{z}i\v{c}kov\'{a}}
\author[7]{Clara Flegel}
\author[7]{Natalia Herrera Valencia}
\author[7]{Will McCutcheon}
\author[7,8]{Mehul Malik\footnote{\href{M.Malik@hw.ac.uk}{M.Malik@hw.ac.uk}} }
\author[4,8,9]{Edgar A. Aguilar}
\affil[1]{\textit{Institute of Physics, Slovak Academy of Sciences, 845 11 Bratislava, Slovakia}}
\affil[2]{\textit{Institute of Computer Science, Masaryk University, 602 00 Brno, Czech Republic}}
\affil[3]{\textit{ICFO-Institut de Ciencies Fotoniques, The Barcelona Institute of Science and Technology, 08860 Castelldefels, Spain}}
\affil[4]{\textit{Institute of Theoretical Physics and Astrophysics,
National Quantum Information Centre, Faculty of Mathematics,
Physics and Informatics, University of Gdansk, 80-952 Gdansk, Poland}}
\affil[5]{\textit{International Centre for Theory of Quantum Technologies, University of Gdansk, 80-308 Gdansk, Poland}}
\affil[6]{\textit{Institute of Science and Technology, 3400 Klosterneuburg, Austria}}
\affil[7]{\textit{Institute of Photonics and Quantum Sciences, Heriot-Watt University, Edinburgh, EH14 4AS, UK}}
\affil[8]{\textit{Institute of Quantum Optics and Quantum Information, Austrian Academy of Sciences, 1090 Vienna, Austria}}
\affil[9]{\textit{AIT Austrian Institute of Technology GmbH, 1210 Vienna, Austria}}

\maketitle

\begin{abstract}
{Our ability to trust that a random number is truly random is essential for fields as diverse as cryptography and fundamental tests of quantum mechanics. Existing solutions both come with drawbacks -- device-independent quantum random number generators (QRNGs) are highly impractical and standard semi-device-independent QRNGs are limited to a specific physical implementation and level of trust. 
Here we propose a new framework for semi-device-independent randomness certification, using a source of trusted vacuum in the form of a signal shutter.
It employs a flexible set of assumptions and levels of trust, allowing it to be applied in a wide range of physical scenarios involving both quantum and classical entropy sources. 
We experimentally demonstrate our protocol with a photonic setup and generate secure random bits under three different assumptions with varying degrees of security and resulting data rates.

}

\end{abstract}

\section{Introduction}

Randomness is an important resource in modern information science. It has a
great number of applications, ranging from randomized sampling, simulations,
randomized algorithms and above all, cryptography. Many of these applications
critically depend on the quality of random numbers, and therefore the design of high quality random number generators (RNGs) is of utmost importance. There are
many different sources of entropy that can be utilized for random
number generator designs. These range from simple to generate but hard to predict
computer data (such as the movement of a mouse cursor on a computer screen or the time
between user keystrokes), to seemingly random physical phenomena (such as
thermal noise or the breakdown in Zener diodes \cite{Somlo75,Stipevic}). In this regard, quantum mechanics offers the possibility of truly random events, such as nuclear decay or photons traveling through a semi-transparent mirror (see \cite{RevModPhys.89.015004} for a review on
quantum random number generators).

The quality of random number generators is traditionally assessed with the
help of statistical tests, or, more recently, machine learning \cite{10.1145/3205455.3205518,8396276}, which can verify that the produced string is
virtually indistinguishable from a truly random string. In essence, however,
such an approach to analyzing random number generators is problematic, because
the statistical tests do not assume anything about the origin of the data they
test. As an example, take the binary expansion of the number $e$   ---   although the
string created in this manner would pass many of the conventionally used
statistical tests, it is obviously not suitable for cryptographic purposes.
This ignorance of the process used to generate the tested random string opens
a window to various security risks. Aside from malicious attacks on the random
number generator, such as inserting back-doors \cite{184527} or displaying a simple bias
towards certain strings \cite{Heninger2012,Lenstra:174943}, its functioning
can be compromised by a simple hardware malfunction, which is often hard to
detect \cite{Barker2012}.

Considerations such as these have recently resulted in a different approach to random number generator designs based on quantum phenomena, where stronger
forms of randomness certificates are possible \cite{Pivoluska2014}. Such
quantum random number generators, introduced in \cite{2009PhDT.......344C} and developed in  \cite{Pironio2010, PhysRevA.87.012335,PhysRevA.87.012336,PhysRevA.90.032313,PLESCH20142938,10.1145/2213977.2213984, doi:10.1137/15M1044333, Bierhorst2018,Liu2018,2018arXiv181013346B}, are called \textit{device-independent}
(DI-RNGs), because they assume very little about the hardware they use. The security proof for these devices is usually based
on Bell-type arguments: the random number generator is composed of several
non-communicating parts and runs a set of randomness-generation rounds, which involve a
predetermined quantum measurement. In a small, randomly chosen fraction of the
run-time, the device is tested. In these \textit{test rounds}, the ability of
the devices to violate Bell-type inequalities is verified \cite{PhysicsPhysiqueFizika.1.195,RevModPhys.86.419}. The violation of local-realism can be seen as a \textit{certificate} that the devices use quantum
measurements and their outcomes are fundamentally unpredictable. 
Since Bell-type arguments do not assume anything about the devices used {apart from space-like separation}, this approach can truly be seen as device-independent. 
The disadvantage of DI-RNGs lies in their implementation   ---   loophole-free Bell
violations have been achieved only recently and under very strict laboratory
conditions \cite{Hensen2015,PhysRevLett.115.250402,PhysRevLett.115.250401}. 

In an attempt to retain the randomness certification capabilities of
DI-RNGs with less stringent experimental
requirements, many \textit{semi-device-independent random number generators}
(SDI-RNGs) have been proposed 
\cite{PhysRevA.94.060301,1367-2630-17-12-125011,PhysRevLett.114.150501,PhysRevA.95.062305,PhysRevA.95.042340,PhysRevApplied.7.054018,1367-2630-17-11-113010,2019arXiv190404819R,2019arXiv190509117V,Xu:16,PhysRevX.6.011020,Avesani2018,2019arXiv190509665D}.
Similar to DI-RNGs, SDI-RNGs include test rounds that are designed to
certify the randomness of their output. However, to make the random number
generators experimentally more feasible, reasonable assumptions about the functioning of some components of the RNG are made, such as a trusted source \cite{PhysRevA.94.060301,1367-2630-17-12-125011,PhysRevLett.114.150501,PhysRevA.95.062305,PhysRevA.95.042340,PhysRevApplied.7.054018,1367-2630-17-11-113010,2019arXiv190404819R} or measurement device \cite{PhysRevX.6.011020,Avesani2018,2019arXiv190509665D}.

{In this paper we present a new approach to semi-device-independent randomness certification that allows for flexible assumptions about the workings of an RNG. }
{What sets our work apart from other SDI-RNG proposals is that our framework is formulated in a high-level abstract language of trusted randomness sources. 
This allows us to certify randomness {in a large number of practical implementations} utilizing both quantum and classical entropy sources. 
Additionally, our framework can work with different levels of trust in particular parts of the RNG, without changing the protocol itself. 
This is in contrast to existing SDI-RNGs, where the protocol relies on a fixed set of assumptions about {specific} parts of the device.} 
{We showcase this flexibility using a photon source and a beam splitter as the source of entropy. Changing the assumptions on the photon source---whether it produces either single photons, coherent/thermal states, or is an unknown source characterized only by its average photon production rate--- is possible in our framework, at the cost of changes in the amount of certifiable entropy. Unlike previous SDI-RNG designs, our implementation therefore comes with a user-defined security/production rate trade-off.} 

The paper is organized as follows. 
In section \ref{sec:GeneralFramework} we introduce our general framework, which consists of three abstract models of entropy sources, and a general protocol to extract perfect randomness from them.
We discuss methods to lower bound the entropy of strings obtained from our protocol in section \ref{sec:results}. 
Section \ref{sec:examples} is devoted to a particular experiment implementing the described entropy sources with the use of a photon source and a beam-splitter. Here we also discuss how different assumptions on the experimental setup change its description within our framework, which results in trade-off between security and randomness production rate.
Finally, in section \ref{sec:experiment} we experimentally implement the 
entropy source described in section \ref{sec:examples} and post-process its outcomes with three different sets of assumptions, based on the amount of trust placed on the photon source.


\section{Results}
\subsection{General framework }

\label{sec:GeneralFramework}
In this section we introduce three different abstract models of randomness, with decreasing level of trust and describe a protocol, which uses a trusted shutter to extract randomness from such sources.

Our basic assumption about the entropy source is that at regular time intervals, it produces a signal with probability $p$, and with probability $1-p$, no signal is produced. Such an assumption on the source is conceptually simple, very natural, and in fact many conventional entropy sources {mentioned above, such as Geiger counters, thermal noise, or the breakdown in Zener diodes} can be modelled in this way. One might argue that such an assumption on the source is too strong, because if one also assumes perfect and trusted signal detectors, extracting randomness from such a source is trivial -- click events can be interpreted as ``1'', and no click events as ``0''. 
Entropy of such an output string is easily calculable and it can be post-processed into a perfectly random string. 
Indeed, early trusted {commercial} quantum random number generators can be
described this way (e.g. IDQuantique \cite{iDQ2019} using a photon source and a beam splitter as an entropy source). 
The main result of this work is that the above assumption on the entropy source can be made sufficient even in the case of partially untrusted measurement device. 

\begin{figure}[ht]
\begin{center}
\includegraphics[]{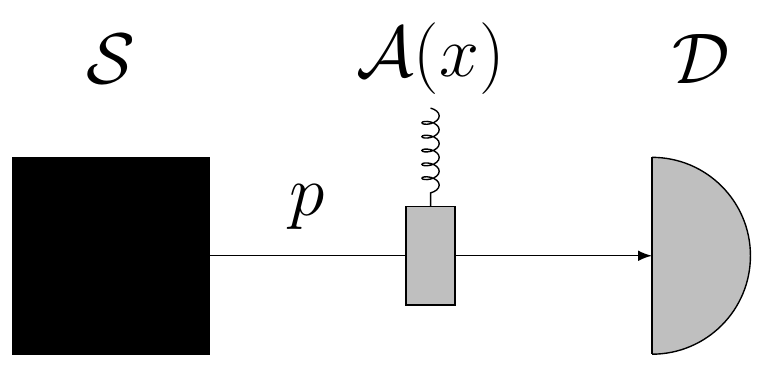}
\end{center}
\caption{The simple scenario consists of a simple entropy source, $\mathcal{S}$,
which emits a random signal with probability $p$. This signal is assumed to be unpredictable to any potential adversary. The signal can be
blocked with the help of a movable shutter, $\mathcal{A}$, controlled via a binary
variable $x$. The measurement device, $\mathcal{D}$, is assumed to be dishonest.}%
\label{fig:device}%
\end{figure}

In order to achieve this, we add an additional component to the setup   ---   a
movable shutter, which can block the signal being sent from the source to the
measurement device (see Fig.~\ref{fig:device}). We call this scenario a \textit{simple
scenario} and the source of entropy a \textit{simple source}.

\begin{figure*}[ht]
\begin{center}
\includegraphics[width = 0.8\textwidth]{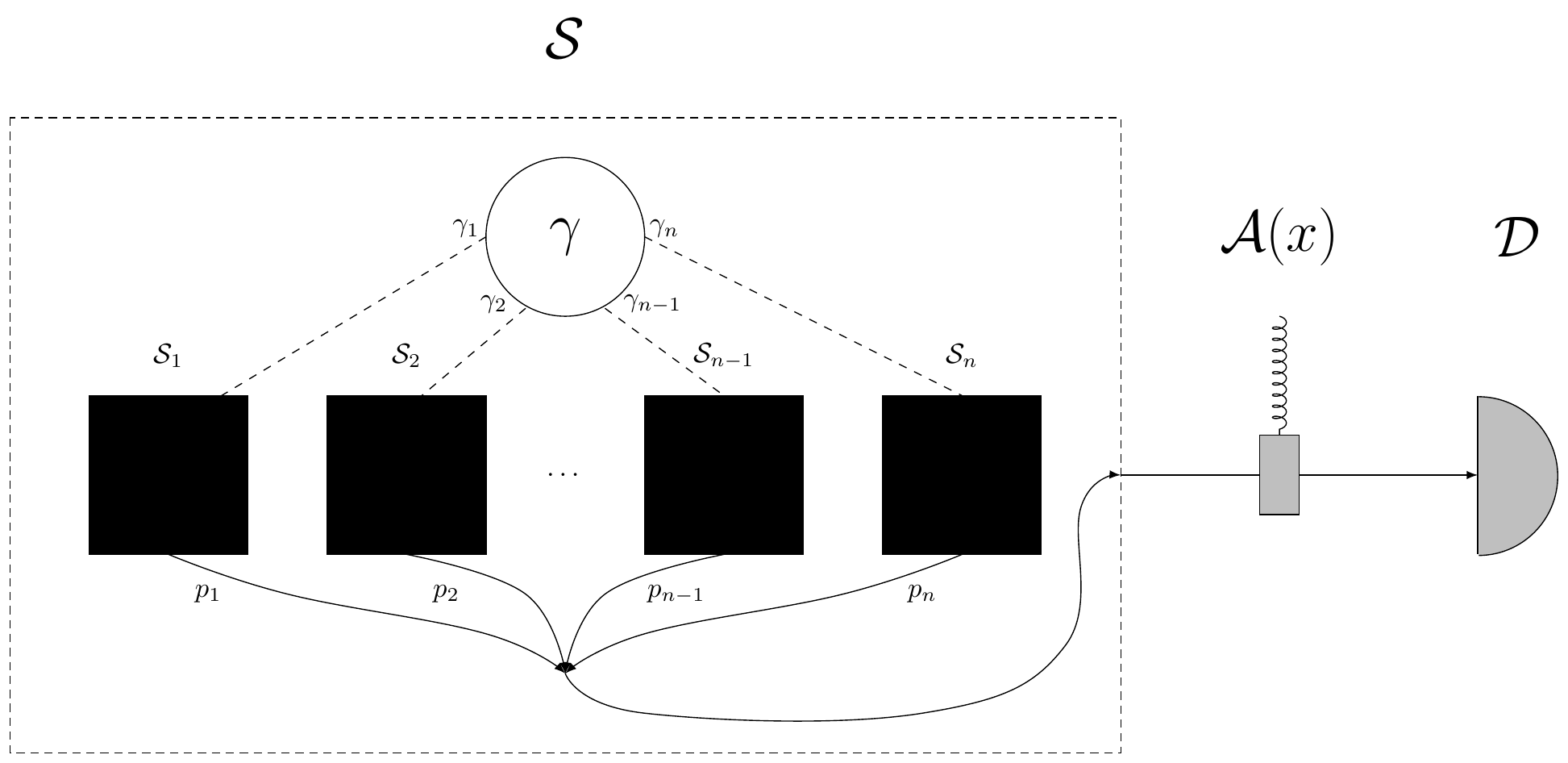}
\end{center}
\caption{ In the second scenario, the entropy source $\mathcal{S}$ (depicted
by a dashed rectangle) is a probabilistic mixture of several (potentially
infinitely many) simple sources $\mathcal{S}_{1},\dots,\mathcal{S}_{n}$.
A random variable $\gamma$ is used to choose a simple source $\mathcal{S}_{i}$,
which emits a random signal with probability $p_{i}$. The choice of source
$\mathcal{S}_{i}$ in a given round is known to the measurement device $\mathcal{D}$ and
the adversary, but not to the user. The random variable $\gamma$ is constrained
either by a fixed probability distribution, or in a more general scenario by more general constraints (e.g. mean value).}%
\label{fig:device3}%
\end{figure*}

Taking the simple entropy source introduced above as a building block, we can generalize to a scenario referred to as a \textit{mixed source scenario}, where the entropy source is a
probabilistic mixture of multiple simple sources. Formally, we define
a discrete (potentially infinite) probability distribution, $\gamma=
\{\gamma_{i}\}, \gamma_{i}\geq 0 \, \forall i, \, \sum_{i} \gamma_{i} = 1$. We associate a simple source $\mathcal{S}_{i}$ with each $\gamma_{i}$.
In the mixed source scenario, the simple source $\mathcal{S}_{i}$ is chosen with
probability $\gamma_{i}$ and subsequently a signal is sent with probability
$p_{i}$ (see Fig.~\ref{fig:device3}).

The value of the random variable $\gamma$ in each round is assumed to be known to the
adversary and the measurement device, but unknown to the user. In order to derive bounds on the entropy produced, the variable $\gamma$ has to be at least partially characterized. This characterization takes the form of a (potentially infinite) sequence of constraints, $\{f_{j}(\gamma) = c_{j}\}$. 
The strongest of such constraint sets is describing $\gamma$ completely by specifying each $\gamma_{i}$. 
We study this special case separately,  
however, we show that the entropy of the RNG output can be lower bounded even for weaker characterizations of $\gamma$. 
In fact, this is possible already if the constraint set contains only a
single (smooth) function (see section~\ref{sec:results}).

Using such a high-level abstraction of the entropy sources is deliberate. 
It makes the extraction protocol presented below usable for a plethora of different experimental setups. 
The only requirement is that trusted randomness is produced in the form of a signal with probability $p$.
The reason why the signal is unpredictable to the adversary varies from implementation to implementation.
This makes the presented framework usable with both quantum sources of randomness, where randomness is guaranteed from inherent non-determinism of certain quantum measurements and classical sources, where some assumption about unavailability of certain data to the adversary must be made.
In the particular experimental realization presented in subsection \ref{sec:examples}, the trusted randomness is obtained from the path a single photon takes travelling through a beam-splitter, which is a genuinely random quantum event.

Before we describe our 
protocol for extracting perfect randomness from the sources described above, 
we list a number of required technical assumptions.

\begin{itemize}
\item The shutter ($\mathcal{A}$) can be reliably controlled by the user
through their inputs $x$.

\item The user has access to a uniform random seed $X$ uncorrelated to the
devices. For example, this can be a private randomness source 
\footnote{In this case we also naturally require that the output randomness  $H_\text{min}(Y|E)$ of the device is longer than $\vert X\vert$. This can be always achieved by increasing the block size $N$ and decreasing the testing rate $q$ accordingly. 
Particularly, the protocol requires $\vert X\vert = N\mathcal{O}(-q\log q)$ bits to choose the test rounds and their shutter settings. With large enough $N$ it is sufficient to set $q = \frac{1}{\sqrt{N}}$, which leads to $\vert X\vert = \mathcal{O}(\sqrt{N}\log{\sqrt{N}})$ bits which are used to choose the test rounds, while a much longer output string of length $H_\text{min}(Y|E) = \mathcal{O}(N)$ is being produced.
Another random string is required for final hashing. This string can however be reused (see section \ref{subsec:Experiment}), thus randomness needed for its selection is negligible for large $N$.}
or public
randomness beacon \cite{NISTBeacon}. 

\item The entropy source $(\mathcal{S})$ is a passive element which does not
change from round to round.

\item The measurement device ($\mathcal{D}$) is memoryless, which together with the previous
assumption implies that each round is identical and independently
distributed.\footnote{The assumption of memoryless measurement devices is rather standard (and often hidden) in the literature, and appears in many contexts (e.g. QKD \cite{McKague_2009}, randomness generation \cite{PhysRevApplied.7.054018}, and more generally, Bell inequality violations \cite{Scheidl2010}). Nevertheless, new methods have recently appeared (e.g.  \cite{8765829} or \cite{PhysRevA.98.040304}), which allow to leave out this assumption.
In particular, using these tools it is often possible to show that the amount of produced private entropy is almost the same as in the memoryless case.}

\item {In case of quantum entropy sources $(\mathcal{S})$ in which the signal state is in a superposition with no-signal state, the measurement device $(\mathcal{D})$ is described by a projector onto a basis that contains the no-signal state (see section \ref{sec:examples} for an example with coherent photon sources).}

\item There is no communication between the devices besides the signal
channel, and the laboratory is shielded from external eavesdroppers. In
particular, neither the measurement device nor the source receive direct information
about the shutter settings $x$.
\end{itemize}

As in any cryptographic protocol, if any of these assumptions cannot be met,
the security of the final string cannot be guaranteed. The assumptions imply that the measurement device is left mostly uncharacterized, in particular, it may still be classically correlated with an adversary.

Now we are ready to present the protocol, which consists of two parts: data
collection and post-processing. For practical purposes, the protocol is run in large batches of $N$ rounds. For the full description, see the box, Protocol 1. As is seen from the protocol, the user will use the testing rounds to obtain a statistical estimation of the workings of the device.
\begin{equation}
\label{eq:ExperimentalS}S = \left(  P(\text{click}|x=0), P(\text{click}%
|x=1) \right)  = (\alpha, \beta)
\end{equation}
More precisely, the user will create a vector $\hat{S}$ as an estimate for $S$ in Eq.~\eqref{eq:ExperimentalS}, which will be filled out with observed experimental
frequencies. This introduces an estimation error $\varepsilon_{e}$, which can be made arbitrarily small by increasing the number of rounds in a batch $N$, and the testing rate $q$. To keep the main text easier to read, we assume that the experimentalist has access to the actual probabilities in Eq.~\eqref{eq:ExperimentalS}, and elaborate on the sampling error in Appendix \ref{appendix:SamplingError}.

\captionsetup[figure]{labelformat=empty}
\begin{figure}
\begin{tcolorbox}[colback=black!2!white,colframe=black,title= Randomness Extraction Protocol]
\textit{Data Collection}
\begin{enumerate}[label = (\arabic*)]
\item In each round  $i \in \{ 1, \ldots, N\}$
 decide whether the current round is a test round ($Q_{i}=\textsc{TEST}$) or a generation round ($Q_{i}=\textsc{GEN}$) at random with probability $(q,1-q)$.
\item If $Q_{i}=\textsc{TEST}$, choose the shutter setting \textit{open}/\textit{closed} ($x_i=0/x_i=1)$), each with probability $\left(\frac{1}{2},\frac{1}{2} \right)$.
Then record the setting and the measurement outcome $(x_i, y_i)$ ($y_i = 1$ if the signal is detected, otherwise $y_i=0$).
\item If $Q_{i} = $\textsc{GEN} 
perform the measurement with the shutter open ($x_{i}=0$) round and record the measurement output $y_{i}$.
\end{enumerate}
\textit{Post-Processing} 
\begin{enumerate}[label = (\arabic*)]
\item Use the rounds $i$ with $Q_{i}=$\textsc{ TEST} to estimate test statistics $S$, (see \eqref{eq:ExperimentalS}).
\item Estimate the min-entropy
$H_\text{min}(Y|E)$  of the random variable $Y = 
\{y_i | Q_i = \textsc{GEN}\}$, (see section \ref{sec:results}). 
\item Choose a security parameter $\varepsilon$ and use a universal hash function on $Y$ to obtain a string $Z$, (see section~\ref{sec:experiment}). 
\end{enumerate}
\end{tcolorbox}
\caption{\textbf{Protocol 1:} An outline of our randomness extraction protocol. For post-processing, note that the estimation depends on the assumptions made on the entropy source used. The output of the protocol $Z$  (of length $H_\text{min}(Y|E)$), is a random variable whose distribution deviates at most $\varepsilon$ in variational distance from a uniform random variable. }
    \label{fig:protocol_box}
\end{figure}
\captionsetup[figure]{labelformat=default}
\setcounter{figure}{2}

In the following section we describe the post-processing procedure. {The goal is to estimate min-entropy $H_\text{min}(Y|E)$ of the output string $Y$ conditioned on the knowledge of the adversary $E$. Min-entropy $H_\text{min}(Y|E)$ roughly describes the length of a perfectly random string obtainable from $Y$ with the help of randomness extractors \cite{10.1007/978-3-642-22012-8_2}.}
The lower bound on min-entropy is obtained by upper bounding the probability of the adversary to guess the outcome of a single randomness generating round (shutter open), denoted $g^*$. 
The obtained upper bound on depends on observed $S$, the type of the entropy source used    ---    simple or mixed, with or without the full characterization of $\gamma$. Finally,  guessing probability, is related to min-entropy of the outcome $Y$ as
\begin{align}\label{eq:guessingToEntropy}
    H_\text{min}(Y|E) \geq -\vert Y\vert \log_2(g^*),
\end{align}
where $\vert Y\vert$ is the number of randomness generating rounds.

\subsection{Entropy estimation}\label{sec:results}

In this section we give a procedure to estimate the entropy of the data collected in the protocol described in the previous section. This is split into three parts based on the type of entropy source used.

\subsubsection{Simple source entropy estimation.}

The simplest case uses a simple entropy source $\mathcal{S}$ which sends a
signal with probability $p$. Based on the assumptions introduced earlier, the strategy of the measurement device to click (i.e. behave as if it detected the signal) in a given round can be based only on
whether the signal arrived or not, (i.e. a single bit of information). 
The response of a detector, (whether to click or not to click) can also be described by a single bit. Therefore, the possible deterministic response functions, called \textit{deterministic detector strategies}, can be described by a function $S:\{0,1\}\mapsto\{0,1\}$, which maps a bit to a bit. 
There are only four functions of this type, which we call
``Never Click" ($S_{N}$) (both input bits are mapped to $0$), ``Always
Click" ($S_{Y}$) (both input bits are mapped to $1$), ``Click Honestly" ($S_{H}$) ($0$ is mapped to $0$ and $1$ is mapped to $1$), ``Click Dishonestly"
($S_{\lnot H}$) ($0$ is mapped to $1$ and $1$ is mapped to $0$).
We represent these strategies by the observable behaviours of the measurement device using them, which can be expressed as vectors
$\left(  P(\text{click}|x=0),P(\text{click}|x=1)\right)  $:%

\begin{equation}
\label{eq:simpleStrategies}%
\begin{split}
S_{N}  &  = \left(  0, 0 \right) \\
S_{Y}  &  = \left(  1, 1 \right) \\
S_{H}  &  = \left(  p, 0 \right) \\
S_{\lnot H}  &  = \left(  1-p, 1 \right)
\end{split}
\end{equation}

General non-deterministic strategies can be described by response functions which, except for the information about the arriving photon, take an additional randomized input. 
It is easy to see that it is sufficient to consider a random input $\lambda$ of $2$ bits, which specifies which of the deterministic functions described above is being used in a given round. 
The observable statistics of such non-deterministic strategies can be therefore
expressed as a convex combination of observable statistics of four deterministic strategies \eqref{eq:simpleStrategies}. The convex combination is described by a hidden variable $\lambda$. We assume that the value of $\lambda$ is shared between the measurement device and the adversary in each round. 
Further, we assume that the values ($Y,N,H$ or $\lnot H$) of the hidden variable $\lambda$
are identically distributed throughout the rounds according to a probability
distribution $\lambda_{Y},\lambda_{N},\lambda_{H},\lambda_{\lnot H}\geq0;
\lambda_{Y}+\lambda_{N}+\lambda_{H}+\lambda_{\lnot H}=1$. 
The adversary tries to guess whether the
measurement device clicked or not in each given round based on their knowledge of $\lambda$, and thus to guess the
outcomes of the RNG. 

Let us highlight the importance of the trusted movable shutter in our design. If the design did not contain it, then setting the random variable $\lambda$ to be uniformly distributed over the strategies $S_{N}$ and $S_{Y}$ would lead to the output of the RNG being uniformly random as well. It would therefore pass any statistical test with high probability,
even though the adversary would posses its perfect copy, rendering it useless for any cryptographic purpose.

In order to safely bound the amount of the entropy produced, the user must assume
that any deviation from the idealized honest scenario $S_{H}$ is correlated
with information gained by the adversary. We measure the information gain by the adversary's optimal guessing
probability, $g^{\ast}$, which is related to the min-entropy 
via $g^{\ast}= 2^{-H_{\text{min}}(Y|E)}$. 
If in a given round the measurement device is following the
strategies $S_{N}$ or $S_{Y}$, then the adversary can be certain of the output, but if $S_{H}$ or $S_{\lnot H}$ is used, 
the guessing probability is
reduced to $g := \max(1-p,p)$.

Without loss of generality, let us assume that $P(\text{click}|x=0) > P(\text{click}|x=1)$ (the other case can be treated similarly, due to the symmetry of the measurement device strategies).
Then, $S_e = (\alpha,\beta)$ can be written as {the convex combination $\frac12(2\beta,2\beta) + \frac12(2\alpha-2\beta,0)$. 
Note that the $(2\beta,2\beta)$ part can be obtained by the measurement device by using only the strategies $S_{N}$ or $S_{Y}$, i.e. without decreasing the adversary's guessing probability. On the other hand, the $(2\alpha-2\beta,0)$ part can be obtained by using strategy $S_H$ only. In particular, this means that whenever our assumption holds, the adversary's optimal strategy is to set $\lambda_{\lnot H} = 0$, and their guessing probability can be obtained by solving the following optimization problem:}%

\begin{align}
\label{eq:singlephoton_LP}g^{\ast}= \max_{\{\lambda\}} \text{ } \text{ }  &
\lambda_{N} + \lambda_{Y} + \lambda_{H} \cdot g\\
\text{s.t.} \text{ } \text{ }  &  \lambda_{Y} + \lambda_{H} \cdot p = \alpha
\nonumber\\
&  \lambda_{Y} = \beta\nonumber\\
&  \lambda_{N} + \lambda_{Y} + \lambda_{H} = 1\nonumber\\
&  \lambda_{N,Y,H} \geq0\nonumber
\end{align}
Since the first three constraints contain only three variables and take the form of equalities, we can directly solve them for 
$\lambda_{\{N,Y,H\}}$ to obtain:

\begin{align*}
\lambda_{Y}=\beta, \quad
\lambda_{H}=\frac{\alpha-\beta}{p}, \quad
\lambda_{N}=1-\beta-\frac{\alpha-\beta}{p}.
\end{align*}
In order to satisfy the last constraint, $\lambda_{N,Y,H}\geq0$, the following needs to hold:
\begin{align*}
0    \leq\beta\leq1\quad
\beta   \leq\alpha\leq\beta+p\quad
\alpha   \leq p+\beta(1-p)\leq\alpha+p.
\end{align*}
These six conditions are not independent, but they can be reduced to three conditions
which are required for the existence of the solution (see Appendix \ref{appendix:Geometric interpretation} for a geometric interpretation):
\begin{equation}
0\leq\beta\leq\alpha\leq p+\beta(1-p). \label{eq:singlephoton_condalpha1}%
\end{equation}
If these conditions are satisfied, the result of the optimization is 
\begin{equation}
g^{\ast}=1-\left(  \alpha-\beta\right)  \left(  \frac{1-g}{p}\right),
\label{eq:singlephoton_solution}%
\end{equation}
{and $H_\text{min}(Y|E)\geq-\log_2\left[  1-\left(  \alpha-\beta\right)  \left(  \frac
{1-g}{p}\right)  \right]|Y|$, where  $|Y|$ is the size of the output string $Y$.}


\subsubsection{Mixed source entropy estimation}\label{subsec:mixedEntropyEstimate}


{Let us now turn to the more involved case of a probabilistic mixture of countably many simple sources.} Recall that in this case the source
$\mathcal{S}$ is a mixture of simple sources $\mathcal{S}_{i}$, characterized
by a known probability distribution $\gamma$. Since the measurement device knows which
source $\mathcal{S}_{i}$ is being used in a given round, it can produce
different statistics $S_{i}$ for each source, and the overall observed statistics can
be written as $S=\sum_{i}\gamma_{i}S_{i}$. Just like in the case of a
single simple source, without loss of generality we assume that $S=(\alpha,\beta)$ satisfies $\alpha\geq\beta$. 
This assumption also implies (see Appendix \ref{appendix:PartialSolutionRestriction}) that in the optimal
solution each $S_{i}=(\alpha_{i},\beta_{i})$ satisfies $\alpha_{i}\geq
\beta_{i},$ as well as the full set of conditions in Eq.~(\ref{eq:singlephoton_condalpha1}). Thus, for each source $\mathcal{S}_{i}$ the
produced statistics can be written as $S_{i}=\lambda_{i,Y}S_{Y}+\lambda
_{i,N}S_{N}+\lambda_{i,H}S_{H_i}$, with $S_{Y}=(1,1)$, $S_{N}=(0,0)$ being the
constant strategies and $S_{H_{i}}=(p_{i},0)$ the honest strategy of the
source $\mathcal{S}_{i}$. Since each source $\mathcal{S}_{i}$ produces entropy {according to}
$g_{i}:=\max(p_{i},1-p_{i})$ and contributes to the overall guessing {probability}
$g^{\ast}$ {by} $g_{i}^{\ast}=\lambda_{i,Y}+\lambda_{i,N}+\lambda
_{i,H} \cdot g_{i}$ {weighted by} $\gamma_{i}$, we have:
\begin{equation}
g^{\ast}=\sum_{i}\gamma_{i}g_{i}^{\ast}.
\end{equation}
Hence, the bound to the adversary's guessing probability is given by the solution to the
following linear program:
\begin{equation}%
\begin{split}
\max_{\{\lambda\}}\text{ }\text{ }  &  \sum_{i}\gamma_{i}(\lambda
_{i,N}+\lambda_{i,Y}+\lambda_{i,H} \cdot g_{i})\\
\text{s.t.}\text{ }\text{ }  &  \sum_{i}\gamma_{i}\left(  \lambda
_{i,Y}+\lambda_{i,H} \cdot p_{i}\right)  =\alpha\\
&  \sum_{i}\gamma_{i} \cdot \lambda_{i,Y}=\beta\\
&  \lambda_{i,N}+\lambda_{i,Y}+\lambda_{i,H}=1\text{ }\forall i\\
&  \lambda_{i,\{N,Y,H\}}\geq0.
\end{split}
\label{eq:knowndistribution_LP}%
\end{equation}
In order to formulate the solution to this optimization problem, let us introduce some notation.
We start by dividing the set of all entropy sources $\mathcal{S}$ into two
sets, $\mathcal{S}_{+}$ and $\mathcal{S}_{-}$. The source $\mathcal{S}_{i}$ belongs to $\mathcal{S}_{+}$ if and only if $p_{i}>\frac{1}{2}$, otherwise
it belongs to $\mathcal{S}_{-}$. Let us also define $N_{+}$ as the number of
sources in the set $\mathcal{S}_{+}$ (including the possibility that $N_{+}$
represents $\infty$). We will use positive integers $i\geq 1$ to label the elements of $\mathcal{S}_{+}$, and negative integers $i\leq -1$ to label the elements of
$\mathcal{S}_{-}$. This allows us to define $N_- = - \vert\s_-\vert$, where $\vert \s_-\vert$ is the cardinality of $\s_-$  (again, potentially infinite). Then, without loss of generality, we will use the
ordering of the sources in the set $\s$ such that
$
\forall i>j,~p_{i}\geq p_{j}.
$
We use the convention that unless specified otherwise, $\sum_{i}$ denotes the sum over all sources from $\mathcal{S}$.
Last but not least, note that we deliberately left out the index $i=0$, as it is used in a formulation of the solution and its proof.

{Using the above notation, the solution of the optimisation problem \eqref{eq:knowndistribution_LP} reads} (see appendix \ref{appendix:KnownDistributionSolution} for proof):
\begin{equation}
g^{\ast}=1-(\alpha-\beta)\left(\frac{1-p_N}{p_N}\right)+\sum_{i=N+1}^{N_{+}}\gamma_{i}
\left(\frac{p_i}{p_N}-1\right).
\label{eq:knownPhotonDistrSolution}%
\end{equation}
Here, if $\sum_{i\in\mathcal{S}_{+}}\gamma_{i}p_{i} \leq \alpha-\beta$, then $ N = 0 $ and $p_N = \frac{1}{2}$, and otherwise $N$ is defined to be the largest natural number such that 
\begin{align}\label{eq:definitionN}
\sum_{i=N}^{N_{+}}\gamma_{i}p_{i}   \geq\alpha-\beta.
\end{align}
{Again, the guessing probability $g^*$ allows us to lower bound the min-entropy of the output string $Y$ of length $|Y|$ as  $H_\text{min}(Y|E) \geq -\vert Y\vert \log_2(g^*)$.}


\subsubsection{Mixed source with partial information}
\label{subsec:PartialInfo}
In the more general case, the probability distribution $\gamma$, which chooses the
simple source $\mathcal{S}_{i}$ to use in a given round, is not fully
characterized, but is constrained by a set of functions, $\{f_{j}(\gamma) = c_{j}\}$. Formally, all the arguments from
the previous case remain the same, except now the optimization needs to be done over the parameters $\gamma_i$ as well.
The maximization task can be stated as follows.
\begin{equation}%
\begin{split}
\max_{\{\lambda\},\{\gamma\}}\text{ }\text{ }  &  \sum_{i}\gamma_{i}%
\left(\lambda_{i,N}+\lambda_{i,Y}+\lambda_{i,H} \cdot g_{i}  \right)\\
\text{s.t.}\quad &  f_j(\gamma)=c_j \quad \forall j\\
&\sum_{i}\gamma_{i}   =1\\
&  \sum_{i}\gamma_{i}\left(  \lambda_{i,Y}+\lambda_{i,H} \cdot p_{i}\right) = \alpha\\
&  \sum_{i}\gamma_{i}\left(  \lambda_{i,Y}+\lambda_{i,H}\right) = \beta\\
&  \lambda_{i,N}+\lambda_{i,Y}+\lambda_{i,H}=1 \quad \forall i\\
&  \lambda_{i,\{N,Y,H\}}\geq0\\
&\gamma_{i}    \geq0.
\end{split}
\label{eq:unknowndistribution_LP}%
\end{equation}
Since the functions $f_j$ are in principle arbitrary, the constraints might not be linear anymore and thus it might not be possible to efficiently solve the problem, even numerically.
However, if the functions $f_j$ are smooth, for every fixed distribution $\gamma$, we are able to optimize over the variables $\{\lambda\}$ according to the previous section. Therefore we can use the solution in Eq.~\eqref{eq:knownPhotonDistrSolution} as the {objective function}, and
the optimization problem becomes:
\begin{equation}%
\begin{split}
\max_{\{\gamma\}}\text{ }\text{ }  &  1-(\alpha-\beta)\left(\frac{1-p_N}{p_N}\right)+\sum_{i=N+1}^{N_{+}}\gamma_{i}
\left(\frac{p_i}{p_N}-1\right),\\
\text{s.t.}\quad &  f_j(\gamma)=c_j \quad \forall j \quad  \\
&\sum_{i}\gamma_{i}   =1\\
&\gamma_{i}    \geq0.
\end{split}
\label{eq:unknowndistribution_LP2}%
\end{equation}
Note that this is still not an easy optimization problem, because 
even if the functions $f_j$ are smooth, a minor change in the distribution of $\gamma_{i}$ might lead to a change in the starting point of the summation in Eq.~\eqref{eq:definitionN}, as $N$ is implicitly dependent on $\gamma_{i}$ via Eq.~\eqref{eq:definitionN}.

To address this problem, let us change the perspective on $N$.  Instead of $N$ being an implicitly defined value dependent on $\gamma$, we will interpret it as a free parameter. Additionally, it can be shown (see section \ref{appendix:KnownDistributionSolution}) that the maximum is obtained when the condition in Eq.~(\ref{eq:definitionN}) for $\gamma$ is satisfied with equality. In such a case 
the objective function can be written in a simpler form (see Eq.~\eqref{eq:knownDistrForMean}) and the optimization problem becomes:

\begin{equation}
\begin{split}
\max_{\{\gamma,N\}}\text{ }\text{ }  &      1- (\alpha - \beta) + \sum_{i=N}^{N^+} {\gamma}_i(2p_i-1),\\
\text{s.t.}\quad & f_j(\gamma)=c_j \quad \forall j \quad\\
&\sum_{i}\gamma_{i}   =1\\
&\gamma_{i}    \geq0\\
&\sum_{i=N}^{N_+}\gamma_{i}p_{i}=(\alpha-\beta).
\end{split}
\label{eq:unknowndistribution_LP3}%
\end{equation}
Note that if we fix the value of $N$, this maximization problem becomes much easier, because the target function is linear in $\gamma$. 
This yields a simple algorithm to find the solution of \eqref{eq:unknowndistribution_LP3}. One can simply solve the problem for each possible $N \in\{1,\dots,N_+\}$, and take the overall maximum over the solutions as the final outcome.

This algorithm of course involves a potentially infinite number of optimization problems to solve, but for simple (e.g.~linear) constraint functions $f_j$ it can be shown that there is a threshold value $N_{\max}$, such that it is not possible to satisfy both the conditions given by $f_j$ and $\sum_{i=N}^{N_{+}}\gamma_{i}p_{i}=(\alpha-\beta)$, whenever $N > N_{\max}$. Last but not least, note that if the functions $f_j$ are linear, for each fixed value of $N$ the optimization problem \eqref{eq:unknowndistribution_LP3} is a linear program and thus can be solved efficiently. 
Additionally, in appendix \ref{appendix:PartialKnowledge} we show that in case of a single linear constraint function $f$, feasible values of $N$ are constrained to a small finite interval, which renders the optimization efficient. 
{The solution to this optimization problem again yields the probability $g^*$ of the adversary to guess the outcome of a single generating round, which is related to min-entropy of output string $Y$ as $H(Y|E)\geq -|Y|\log_2(g^*)$.}
\subsection{Example:  A photon through a beam-splitter}

\label{sec:examples}

\begin{figure*}[ht]
\begin{center}
\includegraphics[width = 0.8\textwidth]{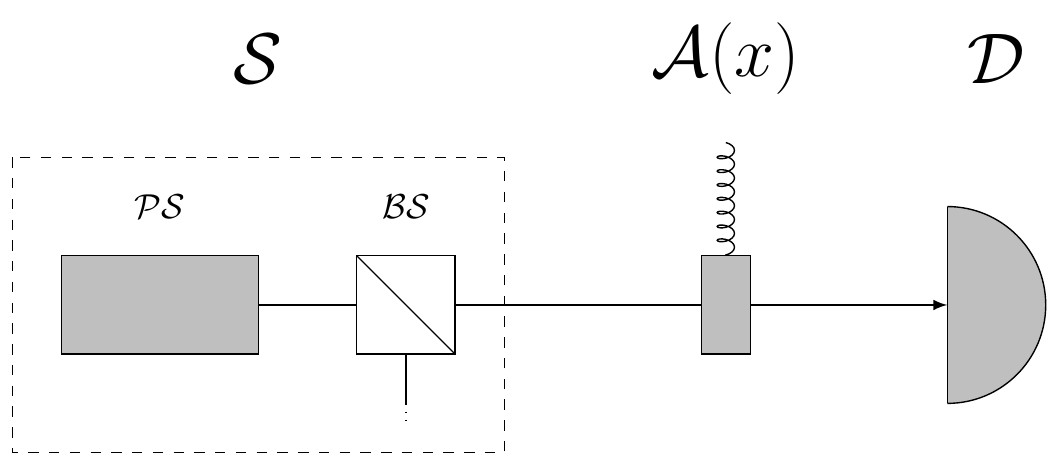}
\end{center}
\caption{The photonic entropy source $\mathcal{S}$ (depicted by a dashed
rectangle) consists of a photon source $\mathcal{PS}$ coupled to a beam-splitter $\mathcal{BS}$ with
the probability of reflection $\pi$ and the probability of
transmission $1-\pi$. Transmitted photons are interpreted as a random signal
emitted from the entropy source, while the reflected photons are discarded. In
order to extract randomness from such a source, we use a mostly uncharacterized and untrusted measurement device
and a trusted shutter controlled by a binary variable $x$. According to the assumptions
we place on the photon source $\mathcal{PS}$, this photonic setup is able to realize all three different general scenarios that we introduced in section~\ref{sec:GeneralFramework}.}%
\label{fig:photonicScenario}%
\end{figure*}

In this section we describe a simple optical setup for randomness generation and analyze it with the help of our framework. 
The entropy source $\mathcal{S}$ consists of a photon source $\mathcal{PS}$ emitting photons
through a beam-splitter $\mathcal{BS}$ with reflection probability $\pi$.
Transmitted photons are coupled to a photon detector $\mathcal{D}$ and their
path can be blocked by a movable shutter $\mathcal{A}$, which can be reliably controlled
via a binary variable $x$. Reflected photons are discarded (see Fig.~\ref{fig:photonicScenario}). 
We use this physical setup to showcase the assumption flexibility our framework allows for.
First of all, the model that describes the entropy source in this setup depends on the assumption we place on the photon source $\mathcal{PS}$. 

\begin{description}
\item[Single photon.] If the photon source $\mathcal{PS}$ produces a single
photon on demand, the entropy source $\mathcal{S}$ is a simple entropy source
with the probability $p = 1-\pi$ of sending a signal.

\item[Known photon distribution.] If the photon source $\mathcal{PS}$ produces
$i$ photons with known probability $\gamma_{i}$, the source $\mathcal{S}$ is a
mixture of simple sources $\mathcal{S}_{i}$ with $p_{i}=1-\pi^{i}$ and
mixing probability $\gamma = \{\gamma_i\}$.

\item[Known mean number of photons.] If the photon source is characterized
only by the mean photon number $\mu$, the setup corresponds to a source
$\mathcal{S}$ which is a mixture of simple sources $\mathcal{S}_{i}$ with
$p_{i} = 1-\pi^{i}$ and the mixing probability $\gamma$ is constrained by
$\sum_{i} i\gamma_{i} = \mu$.
\end{description}

While the single photon source case can be easily seen to be a simple source,
the other two cases require further explanation. Assume that the source
produces an $n$-photon event, where $n\geq2$. Since the number of photons
transmitted through the beam-splitter can vary between $0$ and $n$, the
information available to the (photon-counting) measurement device is more complex than
just binary information about receiving the signal or not. The response
function of the measurement device is therefore potentially more complex than the four
deterministic functions described in Eq.~\eqref{eq:simpleStrategies}.
In fact, there are $2^{n+1}$ different deterministic response functions
assigning click/no-click measurement device events to the number of transmitted photons.
In Appendix \ref{appendix:MultiphotonEventsAreSimpleSources} we show that in
spite of this exponential increase, for each $n$ there are only four response
functions that yield the optimal guessing probability for the adversary. The
first two are ``Never Click" and ``Always Click", which are fully
deterministic and do not depend on the number of received photons. The third
response function is labeled ``Click Honestly". Using this response function,
the measurement device clicks when a positive number of photons arrive and does not
click when no photons arrive. The last response function is called ``Click Dishonestly", and the measurement device clicks only when no photons arrive.
These are exactly the four strategies that characterize a simple entropy
source, since the measurement device decides only on the binary information whether it
received a signal (i.e.~non-zero number of photons) or not. Therefore, known photon distribution case can be characterized as a mixed source with known mixing
probability $\gamma$ and the known mean number of photons case as a mixed source with mixing
probability constrained by mean photon number $\mu$.

{
Note that in the setup described above we do not assume anything about the coherence of the photon source $\mathcal{PS}$.
In fact, in order to be able to describe the strategies available to the measurement device as in the above paragraph, the setup needs to fulfill one out of two assumptions. 
Either $\mathcal{PS}$ produces states which are diagonal in the Fock basis (e.g.\ thermal states),
or the measurement device is measuring in the Fock basis.
In both cases, the mapping to the abstract mixed entropy sources is straightforward.
{Assuming that the state is diagonal in the Fock basis implies} that the whole setup can be implemented with the use of classical sources of light.
On the other hand, the Fock basis measurement assumption is very well motivated from the practical point of view and it allows us some leeway in the description of the $\mathcal{PS}$. %
Namely, we do not require the source to produce a specific photonic state;
it can be characterized solely by a probability distribution $\gamma$ or its mean.
}

Now we are ready to formulate the upper bounds on the guessing probability $g^*$ of a single generating round in case of the known photon distribution and known mean number of photons, {which can be related to min-entropy of the output string $Y$ of the RNG protocol as $H_\text{min}(Y|E)\geq -|Y|\log_2(g^*)$.}

\subsubsection{Known distribution} \label{subsec:KnownDistribution}
{According to the solution of the general case in Eq.~\eqref{eq:knownPhotonDistrSolution},
i}f  $\sum_{i=1}^\infty\gamma_{i}(1-\pi^i)>\alpha-\beta$, we need to find $N$, that is,
the largest natural number such that $\sum_{i=N}^\infty\gamma_{i}(1-\pi^i)\geq\alpha-\beta$. In this case, the optimal guessing
probability is
\begin{equation}
g^{\ast}=1-(\alpha-\beta)\left(\frac{\pi^N}{1-\pi^N}\right)+\sum_{i=N+1}^{\infty}\gamma_{i}
\left(\frac{1-\pi^i}{1-\pi^N}-1\right).
\end{equation}
Otherwise, if $\sum_{i=1}^\infty\gamma_{i}(1-\pi^i)\leq\alpha-\beta$, the optimal guessing probability is
\begin{equation}
g^{\ast}=1-(\alpha-\beta)+\sum_{i=1}^\infty \gamma_{i}(1-2\pi^i).
\end{equation}

\subsubsection{Mean number of photons}\label{subsec:MeanPhoton}

In this section we assume that the photon source is characterized only by its mean photon number $\mu$.
This assumption requires us to solve an optimization
problem of the form as in Eq. \eqref{eq:unknowndistribution_LP3}, where now the condition $f(\gamma) = c_\gamma$ reads
\begin{equation}
\sum_{i=0}^\infty i\gamma_{i}=\mu,
\end{equation}
and $p_i = 1-\pi_i$.

In appendix \ref{appendix:PartialKnowledge} we show that the solution to this optimization problem contains only three non-zero probabilities~$\gamma_i$:
\begin{align}
\gamma_{0} &  =1-\gamma_{N}-\gamma_{N+1}\\
\gamma_{N+1} &  =\frac{\mu-\gamma_NN}{N+1}\\
\gamma_{N} &  =\frac{(N+1)(\alpha-\beta)-(1-\pi^{N+1})\mu}{(N+1)(1-\pi^N)-(1-\pi^{N+1})N}.
\end{align}
for each feasible value of $N=i,~i\in\{1,\dots,N_+\}$.
After plugging these values of $\gamma_0,\gamma_N,\gamma_{N+1}$ into the target function, we obtain the optimal guessing probability: 
\begin{equation}
    g_N^{\ast} = 1 + (\alpha - \beta) - \frac{(\alpha - \beta) + \mu(\pi^{N+1} - \pi^N)}{(N+1)(1-\pi^N) - N(1-\pi^{N+1})}
\end{equation}
The overall solution of \eqref{eq:unknowndistribution_LP3} is $\max_N \{g_{N}^{\ast} \}$ where the the maximization is done over all feasible values of $N$. 
In section \ref{appendix:PartialKnowledge} we show that in general there is only a finite number of feasible values $N$, and therefore the maximum always exists. 
Although the number of these values can still be prohibitively large, in the analysis of the data obtained from the experiment we conducted (see section \ref{sec:experiment}), only a single feasible value of $N$ was encountered, making the analysis very efficient.

\subsubsection{{Other possible assumptions on the photonic setup} }
{
In order to emphasise the flexibility of our framework, in this subsection we discuss possible modifications of the optical setup described above.
Notice that two probability distributions are characterized in the above setup. The first one is the photon number probability distribution $\gamma$ and the second one is the beam-splitter reflection probability $\pi$, or, more generally, binary distributions with probability of success $1-\pi^i$ associated with each photon number. 
The difference between them is that in our setup, the randomness resulting from the beam-splitter events is assumed to be private, unlike $\gamma$, which is available to the adversary. 
Essentially, our framework can be seen as a procedure to certify randomness originating from the trusted source (in this case the beam-splitter) in a noisy setup, where the noise is only partially characterized.}

{
One can, however, assume that the photon emission is also a private random event characterized by $\gamma$. 
This is natural if the photon source $\mathcal{PS}$ is coherent for example, since in this case it is impossible for the adversary to know the photon number in a given round before the measurement.
In such a case, both the entropy originating from the beam-splitter and the entropy of the photon source can be combined into a simple source with the probability of signal $p = 1-\sum_{i=1}^\infty\gamma_i(1-\pi^i)$.
Or, even more interestingly, the beam-splitter can be left out from the setup altogether and assuming Fock measurements, the setup can be analyzed as a simple source with
signal probability $1- \gamma_0$. Note that a similar experiment was studied in two recent works \cite{PhysRevApplied.7.054018,2016arXiv161206828V}, but was analyzed by different techniques.}


\subsection{Experimental realization and results}\label{sec:experiment}
\begin{figure*}[htbp]
    \centering
    \includegraphics[width = 0.8\textwidth]{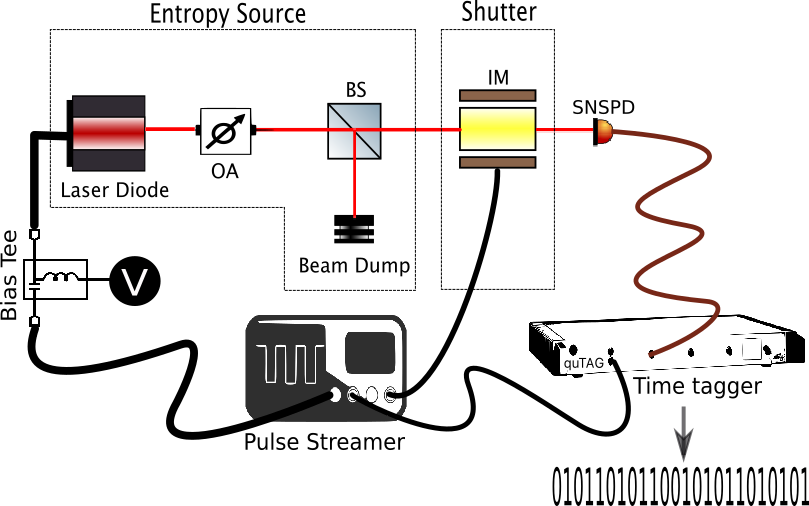}
    \caption{Experimental implementation: The proposed protocols are tested in an optical setup where weak coherent pulses of $8$~ns duration at $5$~MHz are generated by driving a laser diode with a signal generator (Pulse Streamer), and attenuating its power to $\sim1$ photon per pulse. The photons (represented by the solid red line) are incident on a fiber beam-splitter (BS) that discards the reflected photons. The transmitted photons are fast-switched via a fiber electro-optic intensity modulator (IM) that is driven by the digital output of the pulse streamer which provides the uniform random seed}. The pulses at the output of the shutter (composed of the IM) are sent to a superconducting nanowire single photon detector (SNSPD). The counts of the detector and a clock signal from the pulse streamer are recorded with a counting logic (quTAG time-tagger), which allows one to extract coincidences for each pulse and generate the bit string.
    \label{fig:expSetup}
\end{figure*}

{We experimentally implemented the optical random number generation setup described in section \ref{sec:examples}. 
In the experimental implementation (see Fig.~\ref{fig:expSetup}), the photon source $\mathcal{PS}$ is a source of weak coherent pulses, the shutter $\mathcal{A}(x)$ is implemented with an electro-optic intensity modulator (IM)}, and the detection $\mathcal{D}$ is performed by a single-photon detector (SNSPD) and a counting logic (time tagger) to extract the output bit string. The details of the experiment and postprocessing are presented in section \ref{subsec:Experiment}. 

The results of the experiment are summarized in Table \ref{tab:exp_results}. We see that while the actual rates of randomness generation depend on the level of trust into different parts of the experimental setup, even in the most adversarial scenario one can achieve rates comparable to less secure settings. 

\begin{table*}[ht]
\begin{center}
\begin{tabular}{ |c|p{7cm}|c|c|c| } 
 \hline
 \textbf{\#} & \textbf{Assumptions} & \textbf{Cutoff $H_\text{min}$} & \textbf{Batches Used} & \textbf{Extracted Randomness} \\ \hline
 (i) & Single photon source \nl $p_n \sim \delta_{1,n}$  & $0.458$ & $985$ & $37.2$ Mbits \\ 
 (ii) & Poisson photon number distribution \nl $p_n \sim \text{Pois}(\mu)$ & $0.167$ & $975$ & $13.2$ Mbits \\ 
 (iii) & Unknown distribution, mean constraint \nl $\mathbbm{E}[n] = \mu$  & $0.043$ & $948$ & $3.19$ Mbits
 \\ 
 \hline
\end{tabular}
\end{center}
\caption{\label{tab:exp_results} Amount of experimental randomness extracted from different scenarios. In our experiment we have used $\mu = 1.06$. The same raw data ($1,000$ batches consisting of $100,000$ rounds each) was used in each case. For comparison, a hypothetically ideal experiment, with beam splitter transmittance $p=1/2$, single photon source $\mathcal{PS}$, and perfect observed statistics $S = \left(1/2,0\right)$, would produce $69.5$ Mbits from the same data. This takes into account that the data still has to be tested, and the estimator $\hat{S}$ is always taken in a conservative way. Only batches with entropy larger or equal to \textbf{Cutoff $H_{\text{min}}$} were post-processed. Number of such batches for each set of assumptions is stated in column \textbf{Batches used}}.
\end{table*}

\section{Discussion}

{
In this paper, we have presented a novel framework to design and analyze semi-device-independent random number generators.} {In contrast with previous approaches, our framework does not require any fixed assumptions to be made on the workings of an RNG, can be applied in a very broad family of physical implementations, and can cover very different levels of trust placed on different parts of the random number generator.} The centerpiece of our approach consists of a shutter that can trustfully block the transmitted signal. This, in connection with some limited trust in the source and/or measurement devices, proves to be enough to certify randomness.

{During the certification protocol, sample data is collected in order to characterize the behaviour of the measurement devices during both shutter settings (open or closed). 
This sample data is subsequently used to calculate the probability that the adversary, who is classically correlated with the measurement devices, is able to guess the outcome of the measurements with open shutter. 
This calculation involves solving a number of optimization problems expressed as linear programs. The exact formulation and number of these linear programs depends on the level of trust we place into the entropy source. 
Three different trust levels are possible: (i) a simple source emits a signal with probability $p$; or the entropy source is a mixture of multiple such simple sources governed by a probability distribution $\gamma$, which is either (ii) fully characterized; or (iii) partially characterized.
The main benefit of our framework is that all three characterizations can be used with the same physical setup and can be seen as different levels of trust placed onto the entropy source.}

{We showcase the applicability of the framework by implementing a random number generator using a weak coherent optical source and a beam-splitter. This implementation allowed us to demonstrate the novel property of our framework: flexibility in the assumptions made about specific parts of the device. We have data analyzed from a single experiment under three very different sets of assumptions on the source   ---   {true single photons, coherent states, and an unknown source characterized only by its average photon production rate.} In all cases, we were able to extract high-quality random strings, but with significantly different rates. This is natural, as {stronger} assumptions on the source allow for better extraction rates at the cost of giving the adversary more possibilities to attack.}

{Our approach provides significant practical benefits for secure randomness generation. Using the same simple device, a user can make their own choice of the level of secrecy or production rate, just by choosing the appropriate post-processing strategy. Very interestingly, our results show that even for the most adversarial assumption on the source, i.e.~trust only in the mean number of photons, rates of the same order of magnitude were achieved as with the rather strong assumption of a coherent source. The average number of photons produced by a source is testable in principle via its energy consumption, which provides a possible means to further strengthen the security of our framework. Our results pave the way towards practical and experimentally feasible semi-device-independent random number generators, which play a crucial role in the ongoing quantum information revolution.}

\section{Methods}
\subsection{Experimental realization}\label{subsec:Experiment}
In this section we provide details for the experimental realization and post-processing.
Recall that the experiment consists of a source of weak coherent pulses,  electro-optic intensity modulator (IM) implementing the shutter, a single-photon detector (SNSPD) and a counting logic (see Fig. \ref{fig:expSetup}). 
In real world network applications, clock synchronisation of optical pulse trains and detectors would be straightforward. However, for the purposes of this demonstration, all electrical signals are generated by a Swabian instruments Pulse-Streamer $8/2$. 
We drive a laser diode with $0.8$~V analogue pulses of $8$ ns duration (limited by analogue output bandwidth) at $5$~MHz (limited by single-photon detection amplifier deadtime $\sim150$~ns), attenuate the weak coherent pulses to $\sim1$ photon per pulse, and are then incident through a fiber beam-splitter with power transmission $0.5118\pm0.0005$. 
These pulses are fast-switched via a fiber lithium niobite electro-optic intensity modulator (EOSpace, Model AZ-0S5-10-PSA-SFA) driven by a digital output of the signal generator, which with probability  $q=8/100$ blocks the channel (i.e.~TEST rounds with $x=1$). A typical extinction ratio of $1/100$ is observed and a slight thermal drift is calibrated for each $100,000$ rounds of the experiment.
The pulses are then routed to the detectors through a channel with lumped efficiency from switch to detectors of $0.9339\pm0.00005$. 
Detection is made by superconducting nanowire single photon detectors with efficiency $0.9231\pm 0.0007$. 
A QuTools QuTAG counting logic records time-tags from the detectors along with a clock signal from the signal generator, allowing coincidences for each pulse to be extracted using a 10 ns coincidence window, and the output bit string to be recovered. 

Once the data is collected, we begin the post-processing on batches of size $N = 100,000$. With probability $8/92$ we randomly select some of the non-blocked rounds to be TEST rounds with $x=0$ (such that the expected number of test rounds with $x=0$ is the same as the expected number of test rounds with $x=1$). 
Given the large number of total test rounds ($\sim16,000$), we use the Chernoff--Hoeffding bound to calculate the test statistic $\hat{S}$ {(\ref{eq:ExperimentalS})} with sampling error $\eps = 10^{-6}$. In particular, we give a conservative estimate of $S$ and the probability that either $\alpha$ or $\beta$ falls outside of the desired interval is $2\eps$ (see Appendix \ref{appendix:SamplingError} for details). {For each batch, we used $\hat{S}$ to calculate an upper bound on the adversary's guessing probability $g^\ast$. We have performed separate estimations} for the three scenarios; (i) a single photon source, (ii) the photon number distributed according to a Poisson probability distribution with {mean} $\mu$, (iii) the photon number being $\mu$ on average. 
For cases (ii) and (iii), we used $\mu = 1.06$ since it is an upper bound on the observed average photon number per pulse, and that yields the least amount of entropy. 

For simplicity, the length of {the output string} $Y$ was chopped to a constant size of $|Y|=83,000$ per batch, {which leads to a final lower bound on the entropy of the string $Y$, expressed as $H_\text{min}(Y|E) \geq -83,000\cdot\log_2(g^*)$}.  To extract the final random string $Z$ from $Y$ (see the protocol description in Fig.~\ref{fig:protocol_box}), a hashing function, in our case a random binary Toeplitz matrix, has been applied to $Y$.
To keep the discussion clear, we shall focus on case {(ii)} of a known ({Poisson}) distribution. The remaining cases follow an analogous post-processing strategy. In order to reduce the amount of randomness needed, we only generated one Toeplitz matrix and re-used it for every batch. Indeed, the Leftover Hashing Lemma guarantees that when using a Toeplitz matrix of dimensions $|Y|\times(-\log_2(g^*)+2\log\delta)$, the output string $Z$ is at most $\delta$-far in $\ell_1$-distance from being uniformly distributed \cite{Ma_hashing}. We take $\delta = 2^{-100} \approx 10^{-31}$, which implies that we can re-use the Toeplitz matrix $\sim 10^{20}$ times and still maintain a $\ell_1$-distance from the uniform distribution of no more than $10^{-10}$. 

\begin{figure*}
    \centering
    \includegraphics[clip, trim=1.4cm 2.1cm 2cm 1.4cm, width = 0.7\textwidth]{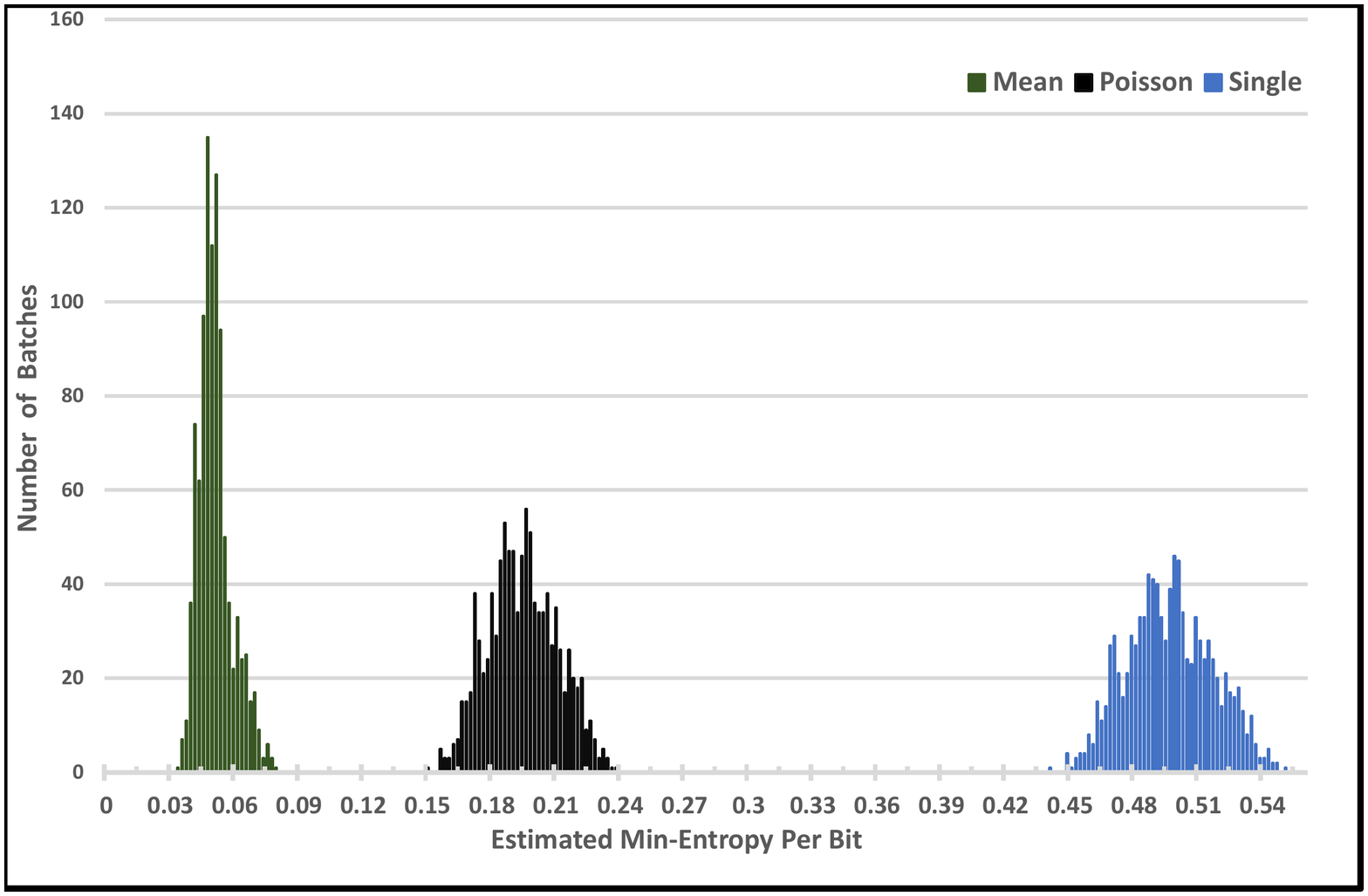}
    \caption{ Distribution of min-entropy estimates depending on the assumptions on the photon source used for $1,000$ experimental batches. From left to right, the assumptions are: (iii) mean photon number $\mu = 1.06$, (ii) Poisson probability distribution with $\mu = 1.06$, and (i) single photon source.}
    \label{fig:experimental_histogram}
\end{figure*}

Note that the estimates of entropy $H_\text{min}(Y|E) \geq -83,000\cdot\log_2(g^*)$ differ for different batches of data. 
This is because  the entropy per bit $-\log_2(g^*)$ is estimated separately for each batch, with different results. The experimental distribution of estimated min-entropies (per bit) can be seen in Fig.~\ref{fig:experimental_histogram}.
However, in order to re-apply the same hash function to each batch, it is important to guarantee that all batches have their min-entropy lower bounded by the same value -- this is because the output length of the hash function must be smaller than the total min-entropy of the input.
If all batches were used, the total min-entropy in each string would have to be bounded by the min-entropy of the worst batch which can be rather low. 
Therefore it is advantageous to discard a (small) number of batches with low certified min-entropy, which increases the amount of min-entropy we can extract per batch, but decreases the number of batches. 
 One can optimize the cutoff threshold min-entropy for each case in order to extract the maximum amount of randomness. 

In the known photon-number distribution scenario {(ii)}, we chose a cutoff of $0.167$ bits of entropy per physical bit, i.e.\ any batch whose estimated min-entropy was lower was simply discarded. Therefore, the Toeplitz matrix generated was of size $83,000 \times 13,661$. For demonstration purposes, we collected data from $1,000$ batches with each batch on average having an estimated $0.185$ bits of entropy per physical bit. $97.5\%$ of the batches were calculated to be above the set threshold, resulting in a total of $13.2$ Mbits of extracted randomness. All results are summarized in Table \ref{tab:exp_results}.

Finally, we carried out the industry-standard NIST randomness tests using an improved implementation presented in \cite{Ss2016Algorithm9O}. As expected, the processed output performed well in all of these tests.

\subsection{Known distribution analysis}

\label{appendix:KnownDistributionSolution}

Here we derive the results presented in subsection (\ref{subsec:mixedEntropyEstimate}).
We need to find the solution to the following optimization problem:

\begin{equation}%
\begin{split}
\max_{\{\lambda\}}\text{ }\text{ }  &  \sum_{i}\gamma_{i}(\lambda
_{i,N}+\lambda_{i,Y}+\lambda_{i,H} \cdot g_{i})\\
\text{s.t.}\text{ }\text{ }  &  \sum_{i}\gamma_{i}\left(  \lambda
_{i,Y}+\lambda_{i,H} \cdot p_{i}\right)  =\alpha\\
&  \sum_{i}\gamma_{i}\lambda_{i,Y}=\beta\\
&  \lambda_{i,N}+\lambda_{i,Y}+\lambda_{i,H}=1\text{ }\forall i\\
&  \lambda_{i,\{N,Y,H\}}\geq0.
\end{split}
\label{eq:knowndistribution_LP2}%
\end{equation}

Recall that we use the following notation:
We start by dividing the set of all sources of entropy $\mathcal{S}$ into two
sets, $\mathcal{S}_{+}$ and $\mathcal{S}_{-}$. The source $\mathcal{S}_{i}$
belongs to $\mathcal{S}_{+}$ if and only if $p_{i}>\frac{1}{2}$, otherwise
it belongs to $\mathcal{S}_{-}$. We define $N_{+}$ as the number of
sources in the set $\mathcal{S}_{+}$ (including the possibility that $N_{+}$
represents $\infty$). We use positive integers $i\geq 1$ for indexing the elements of $\mathcal{S}_{+}$, and negative integers $i\leq -1$ for indexing the elements of
$\mathcal{S}_{-}$. This allows us to define $N_- = - \vert\s_-\vert$, where $\vert \s_-\vert$ is the cardinality of $S_-$ (again, potentially infinite). Then, without loss of generality, we order the sources in the set $S$ such that
\begin{equation}
\forall i>j: ~~ p_{i}\geq p_{j}.\label{eq:ordering}%
\end{equation}
We use a convention that unless specified
otherwise, $\sum_{i}$ denotes the sum through all sources from $\mathcal{S}$.
Last but not least, note that we deliberately left out the index $i=0$, as it is used later in the proof.

Also recall that for the measured parameters to be physical, we require 
\begin{equation}
0\leq\beta\leq\alpha\leq p+\beta(1-p). \label{eq:singlephoton_condalpha}%
\end{equation}
Substituting the equality constrains of Eq.~\eqref{eq:knowndistribution_LP2} together with Eq.~\eqref{eq:ordering}, we can further simplify the optimization problem to%

\begin{align}
\max_{\{\lambda\}}\text{ }\text{ }  &  1-(\alpha-\beta)+\sum_{i\in
\mathcal{S}_{+}}\gamma_{i}\lambda_{i,H}(2p_{i}%
-1)\label{eq:knowndistribution_LP_simpl}\\
\text{s.t.}\text{ }\text{ }  &  \sum_{i}\gamma_{i}\lambda_{i,H} \cdot p_{i}%
=\alpha-\beta\label{eq:knowndistribution_LP_simpl2}\\
& \sum_{i}\gamma_{i}\lambda_{i,Y}   =\beta
\label{eq:knowndistribution_LP_simpl3}\\
& \lambda_{i,N}+\lambda_{i,Y}+\lambda_{i,H}  =1\text{ }\forall
i\label{eq:knowndistribution_LP_simpl4}\\
&  \lambda_{i,\{N,Y,H\}}\geq0. \label{eq:knowndistribution_LP_simpl5}%
\end{align}
It is now easy to see that in order to find the maximum of \eqref{eq:knowndistribution_LP_simpl}, we need to set as many $\lambda_{i,H} = 1$ as possible, starting with ones with the highest parameter $p_i$. 
Of course this needs to be done with the constraints \eqref{eq:knowndistribution_LP_simpl2} -- \eqref{eq:knowndistribution_LP_simpl5} in mind.

Let us start with the simpler of two possibilities.
If
\begin{equation}
\sum_{i\in\mathcal{S}_{+}}\gamma_{i}p_{i}\leq\alpha-\beta,
\label{eq:knowndistribution_Lq}%
\end{equation}
then we can set $\lambda_{i,H}=1$ for all $i\in S_+$ (and therefore $\lambda_{i,N} = \lambda_{i,Y} = 0$ for all $i\in S_+$). 
It remains to show that we can find values for the other $\lambda$ variables such that the solution fulfills all the constraints. 
{Let us first treat the case of both sets $\s_-$ and $\s_+$ being non-empty. The other two special cases will be treated separately later.}
First, let us set for $\forall i\in \s_-$
\begin{equation}
 \lambda_{i,H} = \Delta=\frac{\alpha-\beta-\sum_{i\in\mathcal{S}_{+}}\gamma_{i}p_{i}}%
{\sum_{i\in\mathcal{S}_{-}}\gamma_{i}p_{i}}.
\label{eq:knowndistribution_delta}%
\end{equation}
Let us now show that this is indeed a valid assignment, i.e. $0\leq\Delta\leq1.$ 
Although the positivity of $\Delta$ follows trivially from (\ref{eq:knowndistribution_Lq}), the second inequality is a little more involved. 
In order to show that $\Delta\leq 1$, let us note that since (\ref{eq:singlephoton_condalpha})
holds for each $\alpha_{i}$, $\beta_{i}$ and $p_{i}$ (this is a necessary condition for the statistics produced by $\s_i$ to be physical), due to its linearity it
also holds for $\alpha = \sum_i \alpha_i$, $\beta = \sum_i \beta_i$ and $p=\sum_{i}\gamma_{i}p_{i}$.
Therefore, from (\ref{eq:singlephoton_condalpha}) we get%
\begin{equation}
\alpha-\beta  \leq p-\beta p \leq p = \sum_{i\in\mathcal{S}_{-}}\gamma_{i}%
p_{i}+\sum_{i\in\mathcal{S}_{+}}\gamma_{i}p_{i},
\label{eq:xxx}
\end{equation}
which proves $\Delta\leq1$. Using the values $\lambda_{i,H}=1$ for $i\in S_+$ and $\lambda_{i,H}=\Delta$ for $i\in S_-$, it is straightforward to verify that the
constraint \eqref{eq:knowndistribution_LP_simpl2} is satisfied. 

In order to satisfy
(\ref{eq:knowndistribution_LP_simpl3}) we need to show that 
\begin{equation}\label{eq:condition312}
(1-\Delta)\sum_{i\in\mathcal{S}_{-}}\gamma_{i}\geq \beta.
\end{equation}
Note that because of \eqref{eq:knowndistribution_LP_simpl4}, we have that $\forall i\in \s_-, (1-\Delta) = \lambda_{i,Y}+\lambda_{i,N}$. Therefore,
\begin{equation}
(1-\Delta)\sum_{i\in\mathcal{S}_{-}}\gamma_{i} = \sum_{i\in\mathcal{S}_{-}}\gamma_i(1-\Delta)
= \sum_{i\in\mathcal{S}_{-}}\gamma_i(\lambda_{i,Y}+\lambda_{i,N}).
\end{equation}
If $(1-\Delta)\sum_{i\in\mathcal{S}_{-}}\gamma_{i}\geq\beta$, we can clearly find positive values of $\lambda_{i,Y}$ and $\lambda_{i,N}$, such that $\sum_{i\in\s_-}\gamma_i\lambda_{i,Y} = \beta$
and $\sum_{i\in\s_-}\gamma_i\lambda_{i,N} = (1-\Delta)\sum_{i\in\mathcal{S}_{-}}\gamma_{i} -\beta$ is positive, thus all the constraints of our optimization problem are satisfied.

The first step to prove \eqref{eq:condition312}, is to show that
\begin{equation}\label{eq:necessarySum}
    \sum_{i\in\mathcal{S}_{-}}\gamma_{i}p_{i}    \leq p\sum
_{i\in\mathcal{S}_{-}}\gamma_{i}.
\end{equation}
We have that
\begin{equation}\label{eq:yyy}
    p = \sum_i\gamma_ip_i
    = \sum_{i\in\s_-}\gamma_ip_i+\sum_{i\in\s_+}\gamma_ip_i
    =p_-\sum_{i\in\s_-}\gamma_i + p_+\sum_{i\in\s_+}\gamma_i,
\end{equation}
where $p_- =\frac{\sum_{i\in\s_-}\gamma_ip_i}{\sum_{i\in\s_-}\gamma_i}$, and $p_+ =\frac{\sum_{i\in\s_+}\gamma_ip_i}{\sum_{i\in\s_+}\gamma_i}$. {Notice that $p$ is a convex combination of $p_-$ and $p_+$ with $p_- \le p_+$, and therefore $p_-\leq p\leq p_+$. Then, it also holds that} $\sum_{i\in\s_-}\gamma_ip_i=p_-\sum_{i\in\s_-}\gamma_i\leq p\sum_{i\in\s_-}\gamma_i$.

Now using \eqref{eq:necessarySum} and (\ref{eq:singlephoton_condalpha}) again, we get
\begin{equation}
\beta\sum_{i\in\mathcal{S}_{-}}\gamma_{i}p_{i} \leq\beta p\sum
_{i\in\mathcal{S}_{-}}\gamma_{i}
 \leq(p-\alpha+\beta)\sum_{i\in\mathcal{S}_{-}%
}\gamma_{i}
  = \left(  \sum
_{i\in\mathcal{S}_{-}}\gamma_{i}p_{i}+\sum_{i\in\mathcal{S}_{+}}\gamma
_{i}p_{i}-\alpha+\beta\right)  \sum_{i\in\mathcal{S}_{-}}\gamma_{i}
\end{equation}

\noindent{Since $\s_-$ is non-empty, we have that $\sum_{i\in\mathcal{S}_{-}}\gamma_{i}p_{i} \neq 0$, which leads to}
\begin{equation}
\beta  \leq\left(  1-\frac{-\sum_{i\in\mathcal{S}_{+}}\gamma_{i}p_{i}%
+\alpha-\beta}{\sum_{i\in\mathcal{S}_{-}}\gamma_{i}p_{i}}\right)  \sum
_{i\in\mathcal{S}_{-}}\gamma_{i} = \left(  1-\Delta\right)  \sum_{i\in\mathcal{S}_{-}}\gamma_{i},
\end{equation}
{that is, Eq.~\eqref{eq:condition312} holds, which} proves that it is possible to satisfy all conditions
(\ref{eq:knowndistribution_LP_simpl2}) to
(\ref{eq:knowndistribution_LP_simpl5}) while maximizing the guessing
probability by a suitable choice of $\lambda$'s.
Setting $\lambda_{i,H} = 1, \forall i\in \s_+$ yields
\begin{equation}
g^{\ast}=1-(\alpha-\beta)+\sum_{i\in\mathcal{S}_{+}}\gamma_{i}(2p_{i}-1).
\label{eq:singlephoton_gLq}%
\end{equation}
{Let us now return to the two special cases. First, assume that $\s_+$ is empty.
In such a case \eqref{eq:yyy} is not well-defined (because in the definition of $p_+$ we divide by~$0$). However, the goal of \eqref{eq:yyy} is to prove \eqref{eq:necessarySum}, which in this case holds trivially, since $p = \sum_{i\in\s_-}\gamma_ip_i$ and $\sum_{i\in\s_-}\gamma_i = 1$.}

{It remains to solve the case of $\s_-$ being empty. Then from \eqref{eq:knowndistribution_Lq} we have that $p\leq \alpha-\beta$. Simultaneously, from \eqref{eq:xxx} we have that $p\geq \alpha-\beta$. 
Therefore, $\alpha - \beta = p$, and in order to fulfill \eqref{eq:knowndistribution_LP_simpl2}, we require $\lambda_{i,H} = 1$ for all $i$.
Also, now we can use the identity $\alpha = p+\beta$ and  (\ref{eq:singlephoton_condalpha}) again to derive $p+\beta \leq  p + \beta(1-p)$, which allows a solution only for $\beta = 0$ (otherwise the observed point $(\alpha,\beta)$ is non-physical). 
With $\beta = 0$ it is easy to see that other constrains are satisfied as well and the maximum is equal to $p$. Note that this is in some sense an extreme case, since the observed point such as this can be obtained only with perfectly error-less devices in the limit of the infinite number of rounds.}

Now we deal with the more interesting case of
\begin{equation}
\sum_{i\in\mathcal{S}_{+}}\gamma_{i}p_{i}>\alpha-\beta.
\label{eq:knowndistribution_Hq}%
\end{equation}
In this case we cannot set $\lambda_{i,H} = 1$ for $\forall i \in\s_+$, as this would violate condition \eqref{eq:knowndistribution_LP_simpl2}.
{If we are only concerned with the variables $\lambda_{i,H}$, it is clear that in the optimal case we could set}
%
\begin{equation}
\lambda_{i,H}=0 \quad \forall i\in\mathcal{S}_{-},
\end{equation}
{as sources in $\s_-$ does not contribute to the objective function in Eq.~\eqref{eq:knowndistribution_LP_simpl}.}
Using this we can rewrite (\ref{eq:knowndistribution_LP_simpl}) into%
\begin{equation}
\max_{\{\lambda\}}\text{ }\text{ }1+(\alpha-\beta)-\sum_{i}\gamma_{i}%
\lambda_{i,H} \label{eq:knowndistribution_LP_II}%
\end{equation}
{Next -- still only being concerned with $\lambda_{i,H}$ --} we argue that (\ref{eq:knowndistribution_LP_II}) is maximized by choosing
$\lambda_{i,H}=1$ for the largest $p_{i}$ (large $i$) and zero elsewhere, except for
a single element with  $0<\lambda_{i,H}<1$. This can be seen from the
fact that keeping the sum of $\sum_{i}\gamma_{i}\lambda_{i,H} \cdot p_{i}$ constant
while minimizing $\sum_{i}\gamma_{i}\lambda_{i,H}$ (as it enters the
maximization function with a negative sign) is the same as maximizing
$\sum_{i}\gamma_{i}\lambda_{i,H} \cdot p_{i}$ while keeping $\sum_{i}\gamma
_{i}\lambda_{i,H}$ constant, which is clearly achieved by choosing $\gamma
_{i}\lambda_{i,H}$ large for $p_{i}$ large and vice versa. {In the following, we show that it is indeed possible to choose the values of $\lambda_{i,H}$ according to this procedure, and satisfy all the constraints Eqs.~\eqref{eq:knowndistribution_LP_simpl2} -- \eqref{eq:knowndistribution_LP_simpl5}, by providing an explicit assignment of all the $\lambda$ variables in Eqs.~\eqref{eq:singlephoton_lambda1} -- \eqref{eq:singlephoton_lambda3}.}

{To obtain the explicit assignment, let us first} define the natural number $N$ in the following implicit way%
\begin{align}
\sum_{i=N}^{N_{+}}\gamma_{i}p_{i} &  \geq\alpha-\beta
\label{eq:knowndistribution_N}\\
\sum_{i=N+1}^{N_{+}}\gamma_{i}p_{i} &  <\alpha-\beta
.\label{eq:knowndistribution_N2}%
\end{align}
In case we have that $\sum_{i=N}^{N_{+}}\gamma_{i}p_{i}>\alpha-\beta$, we
perform the following trick: we formally divide the box labeled $N$ into two
boxes, labeled $N-1$ and $N$, both having the same $p_{N}$. The new parameters
$\widetilde{\gamma}_{N}$ and $\widetilde{\gamma}_{N-1}$ will be defined in the
following way%
\begin{align}
\widetilde{\gamma}_{N}&  =\frac{\alpha-\beta-\sum_{i=N+1}^{N_{+}}\gamma_{i}%
p_{i}}{p_N}\label{eq:knowndistribution_gamma_N}\\
\widetilde{\gamma}_{N-1} &  =\gamma_{N}-\widetilde{\gamma}_{N}.
\end{align}
{Note that both values are well defined, because $\gamma_N \in \s_+$, 
and thus $p_N>\frac{1}{2}$.}
All the boxes labeled by $i<N$ are re-labeled $i\rightarrow i- 1$,
utilizing the so-far unused index $i=0$. This new set of boxes will have the
same properties as the old one, it is a mere change of mathematical
description. For this new set it holds that
\begin{equation}
\sum_{i=N}^{N_{+}}\widetilde{\gamma}_{i}p_{i}=\alpha-\beta
.\label{eq:knowndistribution_Neq}%
\end{equation}

\noindent Now we are ready to state the values of the parameters 
$\lambda$ that
maximize $g^{\ast}$ in the following way 
\begin{align}
\lambda_{i,H} &  =1\quad \forall i\geq N\label{eq:singlephoton_lambda1}\\
\lambda_{i,H} &  =0\quad \forall i<N\label{eq:singlephoton_lambda2}\\
\lambda_{i,Y} &  =\omega \quad \forall i<N,\label{eq:singlephoton_lambda3}
\end{align}
with all other parameters given by the condition for their sum. In the above formulas, we use the definition
\begin{equation}
\omega=\frac{\beta}{\sum_{i=N_-}^{N-1}\widetilde{\gamma}_{i}}%
.\label{eq:knowndistribution_omega}%
\end{equation}

\noindent{Note that this is well-defined, as $\sum_{i=N_-}^{N-1}\widetilde{\gamma}_{i} = 0$ would imply that $\s_-$ is empty, as well as $N=1$ and $\widetilde{\gamma_0} = 0$ {(i.e.~the first non-zero $\gamma_i$ is $\gamma_N$, but we can always start the indexing from the first non-zero element, that is, $\gamma_N = \gamma_1$)}. This in turn implies that $\gamma_1 = \widetilde{\gamma_1}$ and \eqref{eq:knowndistribution_Neq} becomes $\sum_{i\in\s_+} \gamma_ip_i = \alpha - \beta$, which  contradicts Eq.~\eqref{eq:knowndistribution_Hq}.}
Now the maximum guessing probability is
\begin{equation}
g^{\ast}=1+(\alpha-\beta)-\sum_{i=N}^{N_{+}}\widetilde{\gamma}_{i}%
.\label{eq:singlephoton_gHq}%
\end{equation}
The only thing that needs to be shown is that $\omega\leq1$, as its positivity
is obvious from its definition (\ref{eq:knowndistribution_omega}). This comes
from the facts that $N\geq 1$, $p\sum_{i=N}^{N_{+}}\widetilde{\gamma}_{i}   \leq\sum_{i=N}^{N_{+}}\widetilde{\gamma}_{i}p_{i}$ (the argument is analogous to \eqref{eq:necessarySum}) in combination
with (\ref{eq:singlephoton_condalpha}):
\begin{equation}
p-p\sum_{i=N_-}^{N-1}\widetilde{\gamma_{i}} = p\sum_{i=N}^{N_{+}}\widetilde{\gamma}_{i}  \leq\sum_{i=N}^{N_{+}}\widetilde{\gamma}_{i}p_{i}
=\alpha-\beta
\leq p-\beta p,
\end{equation}
from which we have that
\begin{equation}
\beta  \leq\sum_{i=N_-}^{N-1}\widetilde{\gamma}_{i},
\end{equation}
and therefore $\omega \le 1$. 

It remains to join (\ref{eq:singlephoton_gLq})
and (\ref{eq:singlephoton_gHq}) into a single formula.
It suffices to plug \eqref{eq:knowndistribution_gamma_N} into (\ref{eq:singlephoton_gHq}) and obtain:
\begin{equation}
g^{\ast}=1-(\alpha-\beta)\left(\frac{1-p_N}{p_N}\right)+\sum_{i=N+1}^{N_{+}}\gamma_{i}
\left(\frac{p_i}{p_N}-1\right).
\end{equation}
Note that if $\sum_{i\in\mathcal{S}_{+}}\gamma_{i}p_{i}>\alpha-\beta$, one needs to calculate $N$ from \eqref{eq:knowndistribution_N} and \eqref{eq:knowndistribution_N2}.
In case $\sum_{i\in\mathcal{S}_{+}}\gamma_{i}p_{i}\leq\alpha-\beta$, we simply set $N=0$ and $p_N = \frac{1}{2}$ and obtain the solution
\eqref{eq:singlephoton_gLq}.

Last but not least, using the modified parameters $\widetilde{\gamma_i}$, the solution can take the following simple form obtained by plugging  \eqref{eq:knowndistribution_Neq} into \eqref{eq:singlephoton_gHq}:
\begin{equation}\label{eq:knownDistrForMean}
    1- (\alpha - \beta) + \sum_{i=N}^{N^+} \widetilde{\gamma}_i(2p_i-1),
\end{equation}
with $N$ explicitly defined by \eqref{eq:knowndistribution_Neq}. This form is particularly useful in the derivation of the case with mixed sources with partially characterized $\gamma$ (see Section \ref{subsec:PartialInfo}), where we can show that in the optimal solution \eqref{eq:knowndistribution_N} holds with equality and therefore $\forall i, ~ \widetilde{\gamma}_i = \gamma_i$.

\subsection{Mean photon number analysis}
\label{appendix:PartialKnowledge}
Here we derive the results presented in subsection \ref{subsec:MeanPhoton}. In subsection \ref{sec:results} we have shown that the optimization problem associated with the scenario with partial information about the mixed entropy source can be stated as 
\begin{equation}
\begin{split}
\max_{\{\gamma\}}\text{ }\text{ }  &  1-(\alpha-\beta)+\sum_{i=N}^{N_{+}}\gamma_{i}(2p_{i}-1),\\
\text{s.t.}\quad & f_j(\gamma)=c_j \quad \forall j\\
&\sum_{i}\gamma_{i}   =1\\
&\gamma_{i}    \geq0\\
&\sum_{i=N}^{N_+}\gamma_{i}p_{i}=(\alpha-\beta).
\end{split}
\label{eq:unknowndistribution_LPapp}%
\end{equation}
We have also argued that the solution to this problem can be obtained by finding the maximum for each fixed value of $N$ in the range $\{1,\dots,N_+\}$, and the overall solution is the largest of these maxima.
In order to proceed with the analytical solution of this problem, let us 
restrict to a single linear constraint function, $c_\gamma = \sum_i a_i\gamma_i$, and 
reformulate the optimization problem using the Lagrange function for each fixed $N$,
\begin{equation}
\mathcal{L}_{N}=1-(\alpha-\beta)+\sum_{i=N}^{N_{+}}\gamma_{i}(2p_{i}%
-1)-\tau_{norm}\left(  \sum_{i}\gamma_{i}-1\right)  -\tau_{f}\left( \sum_ia_i\gamma_i-c_\gamma \right)  -\tau_{N}\left(  \sum_{i=N}^{N_+}%
\gamma_{i}p_{i}-(\alpha-\beta)\right)  , \label{eq:ud_Ln2}%
\end{equation}
with the half-plane conditions $\gamma_{i}\geq0$ for all $i$.

In order to find the maximum, we need to examine 
partial differentiation of (\ref{eq:ud_Ln2}) over all variables $\gamma_i,\tau_{f},\tau_{N},\tau_{norm}$.
While the partial derivatives over $\tau_{f},\tau_{N},\tau_{norm}$ are the required  equality constraints of \eqref{eq:unknowndistribution_LPapp} for $\gamma$, $f$ and $N$, the partial derivatives over all $\gamma_i$ have the following form:%
\begin{align}
\partial_{\gamma_{i}}\mathcal{L}_{N} &  =-\tau_{norm}-\tau_{f}%
a_i \quad \text{if } i < N \label{eq:constraint_i<N}\\
\partial_{\gamma_{i}}\mathcal{L} _{N}&=(2p_{i}-1)-\tau_{norm}-\tau
_{f}a_i-\tau_{N}p_{i} \quad \text{if } i \geq N. \label{eq:constraint_i>N}%
\end{align}

{
Now we need to examine all the stationary points of \eqref{eq:ud_Ln2}. We will argue that on the stationary points, for each variable $\gamma_i$, the corresponding partial derivative is either equal to $0$, or $\gamma_i = 0$ (so that the variable $\gamma_i$ is actually on the boundary of its allowed interval). We first divide the $\{\gamma_i\}$ into two sets: $\Gamma_0 = \{ \gamma_i | \partial_{\gamma_{i}}\mathcal{L}_{N} = 0 \}$ (note that this set cannot contain all the variables, because it is impossible to find values of $\tau_{\{norm,f,N\}}$ for which all partial derivatives \eqref{eq:constraint_i<N} and \eqref{eq:constraint_i>N} vanish), and $\Gamma_b = \{\gamma_i\} \setminus \Gamma_0$. Since by construction the variables in $\Gamma_b$ have non-zero derivatives, by the extreme value theorem the maximum of \eqref{eq:ud_Ln2} must be attained when these variables are on their boundary. This is when $\gamma_i = 0$, since all other constraints are taken care of with the derivatives $\partial_{\{\tau_f, \tau_N, \tau_{norm}\}}\mathcal{L}_{N} = 0$. A maximum may therefore be found if for all $\gamma_i \in \Gamma_b$ we have $\partial_{\gamma_{i}}\mathcal{L}_{N} <0$ (i.e.~the value of $\mathcal{L}$ is increasing towards the boundary of all $\gamma_i \in \Gamma_b$), and since $\mathcal{L}_N$ is linear in all $\gamma_i$ this maximum would be a global one. The remaining issue is therefore to find the optimal set $\Gamma_0$ for which the derivatives \eqref{eq:constraint_i<N} and \eqref{eq:constraint_i>N} vanish. We proceed to construct this optimal set by showing how alternative choices cannot be the optimal solution.
}

Further analysis now depends on the exact values of $a_i$ and $p_i$.
In section \ref{sec:experiment} we have shown that in our experiment, if we characterize the photon source with the mean number of photons only, the constraint function is $\mu = \sum_i i\gamma_i$. 
{ Therefore, we will focus on the case where $\{a_i\}$ is a non-negative, unbounded, and strictly increasing sequence, with our prime example being $\{a_i = i\}$. Likewise $\{p_i = 1 - \pi^i\}$ in our experimental section, so we require $\{p_i\}$ to be a non-negative strictly increasing sequence such that $\{p_i/a_i\}$ is strictly decreasing.
Since the sequence $\{a_i\}$ is unbounded, we need to have $\tau_{f}\geq0$,
otherwise both (\ref{eq:constraint_i<N}) and (\ref{eq:constraint_i>N}) will become positive for some (large enough) value of $i$.}
One (trivial) solution is choosing $\tau_{f}=0$, which is only possible for
$\tau_{norm}\geq0$ {by \eqref{eq:constraint_i<N}}. This would allow to have all $\gamma_{i}$ for $i<N$
potentially non-zero. But then (\ref{eq:constraint_i>N}) will become

{
\begin{equation}
\partial_{\gamma_{i}}\mathcal{L}_{N}=\left(  2-\tau_{N}\right)p_i
-1-\tau_{norm}.
\end{equation}
}
{
If $\tau_{N}\geq2$, then this equation is always negative, leading to  $\gamma_{i}=0$ for
$i\geq N$, which would violate the last constraint \eqref{eq:unknowndistribution_LPapp}. For smaller
$\tau_{N}$, assume $\partial_{\gamma_{i}}\mathcal{L} _{N}=0$ for some $i$.
Then, since the $\{p_i\}$ are increasing, $\partial_{\gamma_{i+1}}\mathcal{L} _{N}>0$. As we have argued above, that would not lead to a maximum, since $\gamma_i\geq 0$, and the derivatives of the $\Gamma_b$ variables should be negative.  Therefore we conclude that $\tau_{f}>0$.}

{  For $i<N$, Eq.~\eqref{eq:constraint_i<N} now reads $\tau_{norm} = -\tau_f a_i$. Since all of the $\{a_i\}$ are different, this equation can only be satisfied for a single variable $\gamma_i$. Notice however that \eqref{eq:constraint_i<N} is a decreasing function in $i$, therefore in order to guarantee that all the non-zero partial derivatives $\partial_{\gamma_i} \mathcal{L}_N$ are negative, we must have $\tau_{norm} = -a_0 \tau_f$. That is, $\gamma_0\in \Gamma_0$ and $\gamma_{0<i<N}\in \Gamma_b$, i.e.~$\gamma_{i}=0$ for $0<i<N$.  }
Eq.~\eqref{eq:constraint_i>N} now reads:
{
\begin{equation}
i\geq N:\partial_{\gamma_{i}}\mathcal{L} _{N}= (2-\tau_{N})p_i - (a_i - a_0) \tau_{f}-1.
\label{eq:constraint_with_a0}
\end{equation}
}
Since we have two free parameters available ($\tau_{f}$ and $\tau_N$), it is possible to achieve $\partial_{\gamma_{i}}\mathcal{L} _{N}=0$ for at most two
different values of~$i$. {Notice that $(a_i-a_0)>0$, and we have shown that $\tau_f>0$.} Therefore $(2-\tau_{N})$ must be positive, { or else all $\partial_{\gamma_{i\geq N}}\mathcal{L} _{N} <0$. Furthermore, since $\{p_i/a_i\}$ is strictly decreasing, then \eqref{eq:constraint_with_a0} is also strictly decreasing. Therefore, in order to satisfy \eqref{eq:constraint_i>N} for two different $\gamma_i$ and to have all the rest of the partial derivatives negative, it must hold that $\partial_{\gamma_{i= N}}\mathcal{L} _{N}=0$ and $\partial_{\gamma_{i= N}}\mathcal{L} _{N+1}=0$. The conditions cannot be solved for the rest, so $\gamma_{i>N+1} = 0$}.

{
 Now, we know that $\Gamma_0 = \{\gamma_0, \gamma_N, \gamma_{N+1}\}$ are the only non-zero variables. We can therefore use the original problem constraints to solve for the unknowns. Namely:
}

{
\begin{align}
1 &= \gamma_0 + \gamma_{N} + \gamma_{N+1}\\
c_\gamma &= a_0 \gamma_0 + a_N \gamma_N + a_{N+1}\gamma_{N+1} \\
\alpha - \beta &= p_N \gamma_N + p_{N+1}\gamma_{N+1}
\end{align}
}
{ This linear system of equations is then solved. The only difficulty remaining is that, depending on the values of $\{a_i\}$ and $\{p_i\}$, it is not at all clear that the solutions satisfy $\gamma_i\geq 0$ for a given $N$. In fact, we will show that in our prime example, $\{a_i = i\}$, $c_\gamma = \mu$, and $\{p_i = 1 - \pi^i\}$, only a finite number of $N$ can satisfy the positivity constraints for $\gamma$. We therefore switch to this concrete example to finish this section. The solution to the linear system of equations reads:
}

\begin{align}
\gamma_{0} &  =1-\gamma_{N}-\gamma_{N+1}\\
\gamma_{N+1} &  =\frac{\mu-N \gamma_N}{N+1}\\
\gamma_{N} &  =\frac{(N+1)(\alpha-\beta)-\mu (1-\pi^{N+1})}{(N+1)(1-\pi^N)-N(1-\pi^{N+1})}.
\end{align}
Note that $\gamma_N$ is approaching infinity with increasing $N$. This means that only a finite number of values $N$ need to be tested, as for sufficiently large $N$ we have $\gamma_{N}>1$ and the positivity constraints for $\gamma_{N+1}$ and $\gamma_0$  cannot be satisfied.
Therefore, the final guessing probability will be the maximum from the \textit{finite} number of guessing probabilities of the form:

{
\begin{equation}
    g_N^{\ast} = 1 + (\alpha - \beta) - \frac{(\alpha - \beta) + \mu(\pi^{N+1} - \pi^N)}{(N+1)(1-\pi^N) - N(1-\pi^{N+1})}.
\end{equation}
}


\section*{Data availability}
The data that support the findings of this study are available from the corresponding author upon reasonable request. 

\section*{Author contributions}
EAA, MF, MPi and MPl formulated the initial idea, EAA, MPi, MPl and NR developed the theory, CF, NHV, WMC and MM performed the experiment, EAA, MPi and MPl analyzed data. All co-authors contributed to the preparation of the manuscript.

\section*{Competing interests}
Authors declare no competing interests.

\section*{Acknowledgements}
We would like to thank Robert Fickler for discussions about the experimental realization and Marek S\'{y}s for running the NIST randomness test on the data we acquired in the experiment. We would like to thank Ugo Zanforlin, Gerald Buller, Daniel White, and Cristian Bonato for their help with the experiment. MPi, MPl, and MM acknowledge Czech-Austrian project MultiQUEST (I 3053-N27 and GF17-33780L). MPi and MPl additionally acknowledge the support of VEGA project 2/0136/19. MF acknowledges support from the Polish NCN grant Sonata UMO-2014/14/E/ST2/00020, the European Research Council (ERC) under the European Union’s Horizon 2020 research and innovation programme ERC AdG CERQUTE (grant agreement No 834266), the State Research Agency (AEI) TRANQI (PID2019-106888GB-I00 / 10.13039/501100011033), the Government of Spain (FIS2020-TRANQI; Severo Ochoa CEX2019-000910-S), Fundació Cellex, Fundació Mir-Puig, and Generalitat de Catalunya (CERCA, AGAUR). MM, WM, NHV, and CF acknowledge support from the QuantERA ERA-NET Co-fund (FWF Project I3773-N36) and the UK Engineering and Physical Sciences Research Council (EPSRC) (EP/P024114/1).

\bibliographystyle{ieeetr}
\bibliography{STRNG}{} 

\appendix
\appendixpage
\section{Geometric interpretation of the optimization problems}

\label{appendix:Geometric interpretation}

\begin{figure}[ht]
\centering
\includegraphics[scale = 0.7]{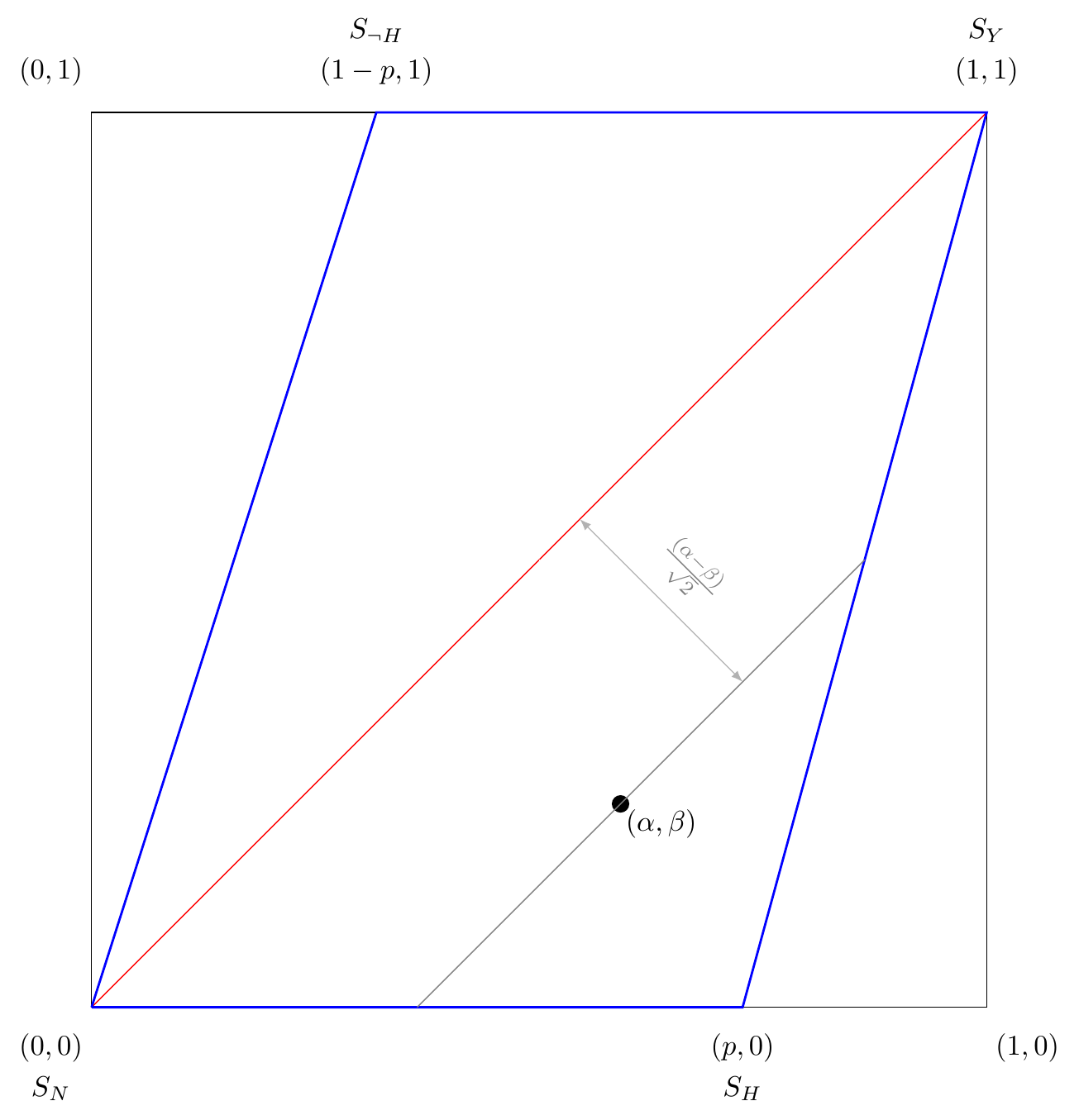} \caption{
Since observed probabilities $S = (\alpha,\beta)$ are mixtures of deterministic strategies, the
space of possible values of $S$ can be depicted as the blue polytope. With
the assumption that $\alpha\geq\beta$, i.e.~the measurement devices click more often
when the shutter $\mathcal{A}(x)$ is open, we can restrict the adversary to
use of three strategies $S_{Y},S_{N},S_{H}$ only. In such a case, the
decomposition of the observed probabilities $S = (\alpha,\beta)$ into
deterministic strategies is unique.  }%
\label{fig:simpleSourcePolytope}%
\end{figure}

In this section we present a geometric interpretation of the problems 
of finding $g^{\ast}$ for both a simple source and a mixed source with fixed $\gamma$.
The geometric interpretation is helpful to build an intuition which in turn helps understanding the formal solutions in the main text.
Let us start with a simple source $\mathcal{S}$ with probability $p$ of emitting a random signal.
Recall that the problem can be phrased as the following
linear program:
\begin{align*}
g^{\ast}=\max_{\{\lambda\}}\text{ }\text{ }  &  \lambda_{N}+\lambda
_{Y}+\lambda_{H} \cdot g\\
\text{s.t.}\text{ }\text{ }  &  S=\lambda_{N}S_{N}+\lambda_{Y}%
S_{Y}+\lambda_{H}S_{H}\\
&  \lambda_{N}+\lambda_{Y}+\lambda_{H}=1\\
&  \lambda_{N,Y,H}\geq0,
\end{align*}
where $g = \max \left( p, 1-p \right)$ is the probability to guess the outcome of the source, and $S=\left(  P(\text{click}|x=0),P(\text{click}|x=1)\right)$ is the
description of the device behaviour estimated during the run of the protocol. Likewise, 
\begin{align*}
S_{N}  &  =(0,0)\\
S_{Y}  &  =(1,1)\\
S_{H}  &  =(p,0)\\
S_{\lnot H}  &  =(1-p,1),
\end{align*}
are the possible deterministic behaviours of the measurement device.
As we have shown in the main text, conditioned on
\[
0\leq\beta\leq\alpha\leq p+\beta(1-p),
\] the solution to this problem is
\begin{equation}\label{eq:appAgast}
g^{\ast}=1-\left(  \alpha-\beta\right)  \left(  \frac{1-g}{p}\right).
\end{equation}
In fact, this has a  {simple} geometric interpretation shown in Fig.~\ref{fig:simpleSourcePolytope}.

{All possible observed statistics are convex combinations of the points $S_N$, $S_Y$, $S_H$ and $S_{\lnot H}$ (inside the blue polytope in Fig.~\ref{fig:simpleSourcePolytope}).}
Note that the deterministic strategies $S_{N}$ and $S_{Y}$ allow the measurement devices to
obtain only the observed probabilities lying on the line $\alpha=\beta$ (red diagonal line in Fig.~\ref{fig:simpleSourcePolytope}). The strategies represented by these points have
guessing probability~$1$. On the other hand, the point $(p,0)$ represents the
honest strategy $S_{H}$, which has guessing probability $g=\max(p,1-p)$. 
It follows that the observed points with the same distance $\frac{(\alpha-\beta)}{\sqrt{2}}$ from the diagonal line $\alpha=\beta$  (depicted as the gray line in Fig.~\ref{fig:simpleSourcePolytope}) have the same guessing probability, as they are convex combinations of strategies on the diagonal and the honest strategy, with the same weight on the strategies on the diagonal. Additionally, the smallest guessing probability (and thus the highest certified entropy) is obtained for the point $S_H$, which is the point farthest from the {$\alpha = \beta$}  line. 
{A crucial fact, that is important to understand how the optimal solution 
in the case of mixed entropy sources looks like, is that when $p\leq \frac{1}{2}$, the {optimal} guessing probability {$g^\ast$ in Eq.~\eqref{eq:appAgast}} does not depend on $p$ at all, as in this case $g = 1 - p$. Then we have $g^* = 1- (\alpha-\beta)$, that is, the optimal guessing probability decreases with a rate proportional to the distance from the diagonal.
In other words, an observed point $(\alpha,\beta)$ certifies the same amount of entropy for all simple sources with $p\leq\frac{1}{2}$.
On the other hand, if $p>\frac{1}{2}$, the rate at which the guessing probability decreases with the distance from the $\alpha = \beta$ line depends on $p$ as $\frac{(1-p)}{p}$, which is a decreasing function of $p$. That is, an observed point $(\alpha,\beta)$ certifies more entropy for sources with a smaller parameter $p$.
Additionally, it is important to note that the conditions for the solution $0\leq\beta\leq\alpha\leq p+\beta(1-p)$ correspond to $(\alpha,\beta)$ being in the blue polytope, below the $\alpha=\beta$  line {(that is, the observed statistics is feasible and $\beta \leq \alpha$)}. }

\begin{figure}[ht]
\centering
\includegraphics[scale = 0.49]{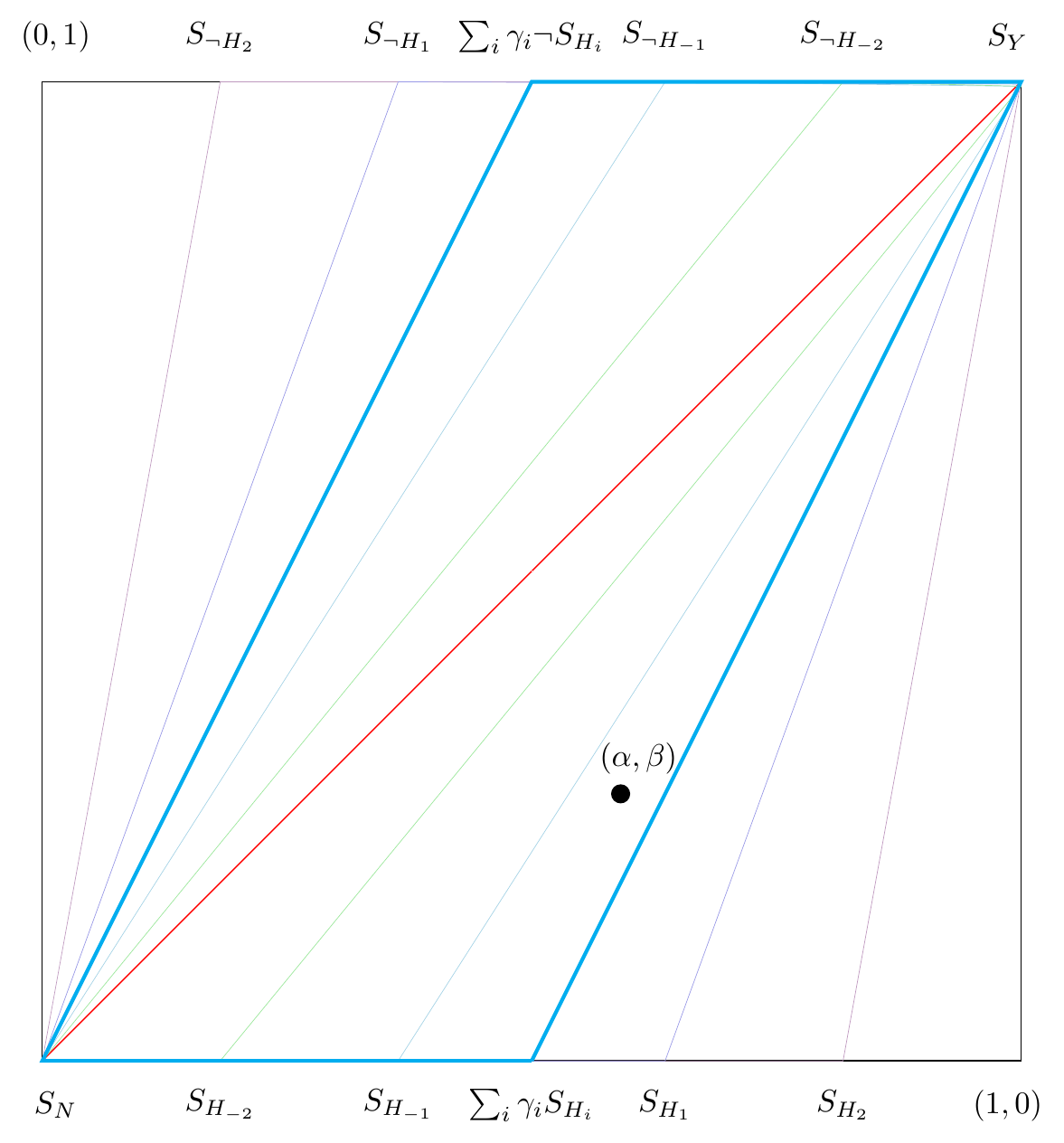}
\includegraphics[scale = 0.49]{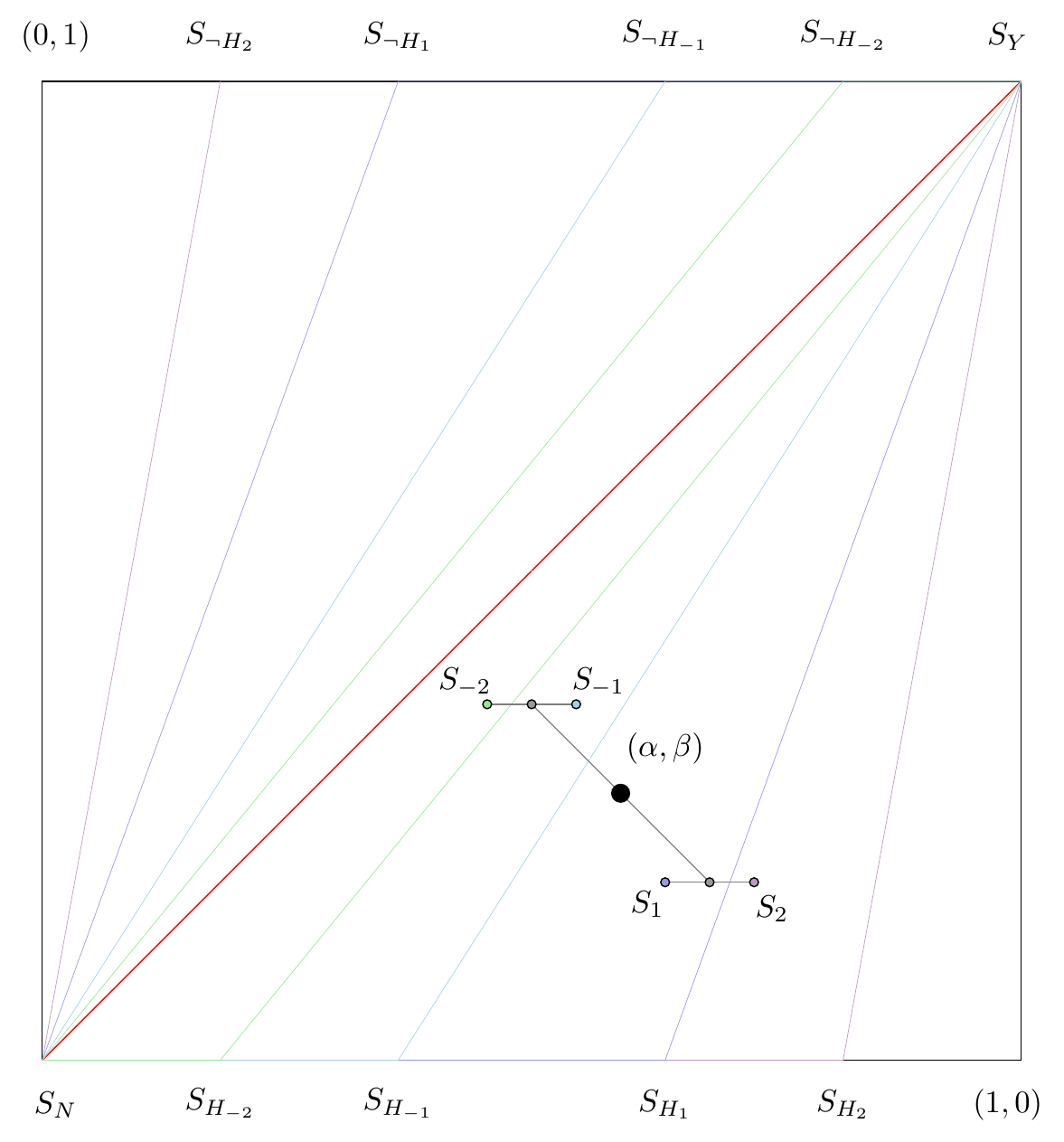}
\includegraphics[scale = 0.49]{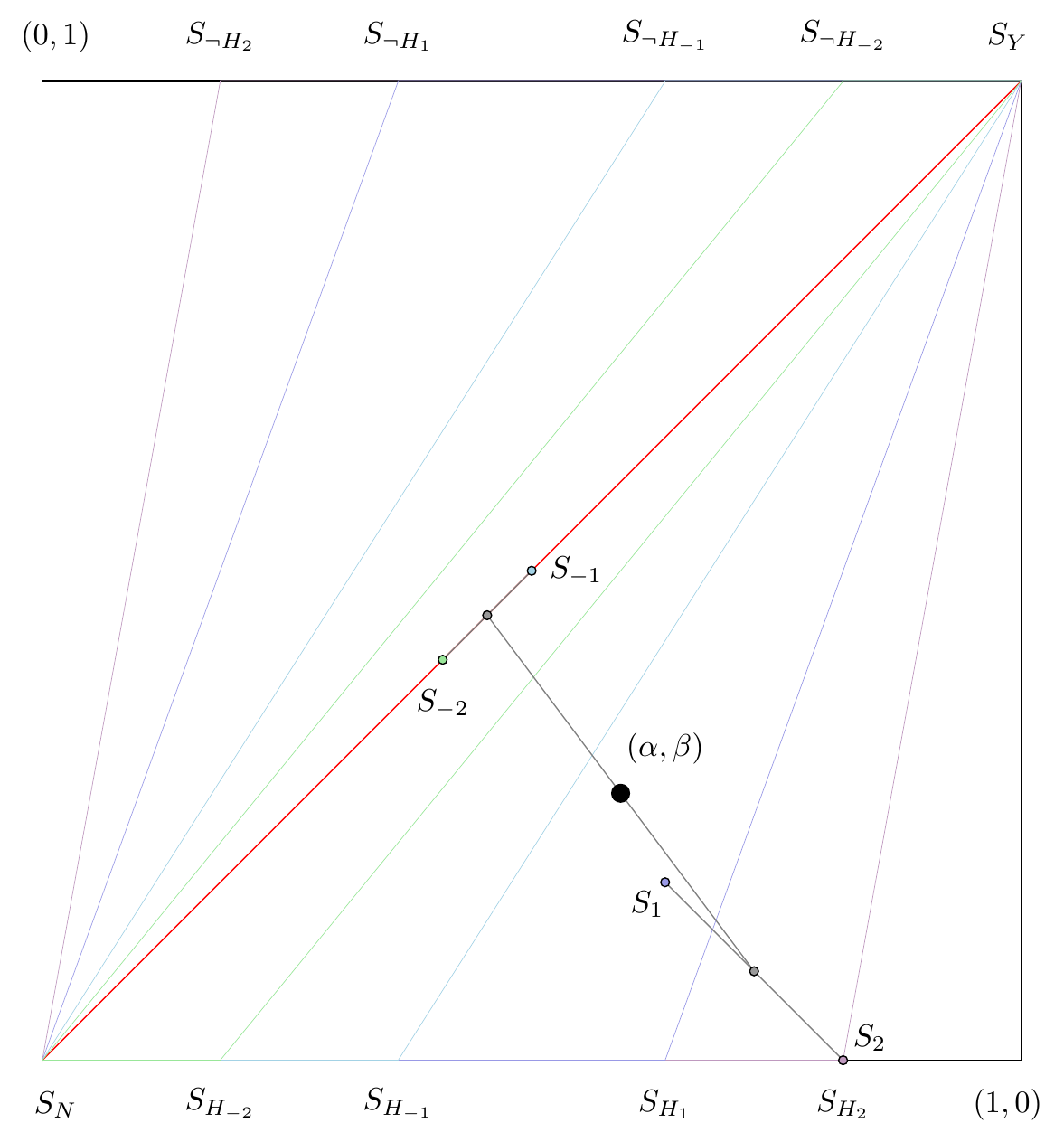}
\caption{{
Here we provide an example with a mixed source $\s$ being a mixture of four sources $S_{-1},S_{-2},S_1,S_2$, with $\gamma_{-2} = \gamma_{-1} = \gamma_{1}=\gamma_2 = \frac{1}{4}$. 
Note that simple sources with $p_i\leq \frac{1}{2}$ have negative index and belong to $\s_-$ and sources with $p_i > \frac{1}{2}$ have positive index and belong to $\s_+$.
In the left figure we depict a (cyan) polytope to which all possible observed points $S$ are constrained.
It can be seen as a weighted average of the polytopes constraining each simple source in the mixture.
In the middle figure we show a feasible (sub-optimal) solution of the optimization problem to maximize the adversary's guessing probability.
The observed  point $S = (\alpha,\beta)$ is a weighted mixture of the points $S_{-2},S_{-1},S_{1},S_2$, which are constrained into their corresponding polytopes defined by the strategies $S_{\lnot H_i}$ and $S_{H_i}$. 
In the right figure we show an optimal solution with $S_{-1}$ and $S_{-2}$ on the zero-entropy diagonal (i.e.~$\lambda_{-2,H} = \lambda_{-1,H} = 0$), $S_2 = S_{H_2}$ (i.e.~$\lambda_{2,H} = 1$) and $S_1$ an intermediate point. Note that only $S_1$ and $S_2$ contribute to the distance of $(\alpha,\beta)$ from the diagonal. 
}}%
\label{fig:mixedSourcePolytope}%
\end{figure}

{
The geometric interpretation of a mixed source is slightly more involved. 
Feasible observed points $S = (\alpha,\beta) = \sum_i \gamma_i S_i$ are mixtures (according to a probability distribution $\gamma$) of feasible points $S_i$ of sources $\s_i$.
The feasibility of $S_i$ {means} that it is constrained into its corresponding polytope defined by the strategies $S_{\lnot H_i}$ and $S_{H_i}$ (see Fig.~\ref{fig:mixedSourcePolytope} for an example with a mixture of four simple sources).
This implies that {all} feasible points $(\alpha,\beta)$ are constrained into a polytope defined by the deterministic strategies $S_N,S_Y$ and the weighted averages of the non-deterministic strategies $\sum_i \gamma_i S_{H_i} = (\sum_{i} \gamma_i p_i,0)$ and $\sum_i \gamma_i \lnot S_{H_i} = (1-\sum_{i} \gamma_i p_i,1)$ {(the cyan polytope on the left subfigure of Fig.~\ref{fig:mixedSourcePolytope})}}

{

{Every non-extremal point of the feasible polytope can be decomposed into convex combinations of statistics of the simple sources, $S_i$,} in infinitely many ways.
Let us now argue, that the decompositions which lead to the highest guessing entropy (and therefore are optimal), have a specific form. There are essentially three different cases.}

{
Let us first deal with the easiest case, in which all the simple sources in the mixture belong to $\s_-$, i.e.~$p_i \leq \frac{1}{2}$ for all~$i$.
As we have discussed above,
in such a case the contribution of each simple source $\s_i$  to the total guessing probability depends only on the distance of $S_i$ from the diagonal, and not on the value of $p_i$.
Note that for any linear decomposition of $(\alpha,\beta)$ into $S_i$ with weights $\gamma_i$, the weighted distance of the points $S_i$ from the diagonal is equal to $\frac{\alpha-\beta}{\sqrt{2}}$. Therefore, each decomposition leads to the same guessing probability $1-(\alpha-\beta)$.}

{
Whenever the mixed source contains also sources with $p_i > \frac{1}{2}$ (i.e.~$\s_+$ is non-empty), the problem becomes more interesting. 
Again, the weighted distance of all the points $S_i\in S$ from the diagonal is constant, however, the guessing probability contribution of the sources in $\s_+$ depends both on the distance from the diagonal and their parameter $p_i$ -- the larger the parameter $p_i$, the smaller the contribution per distance from the diagonal. This simple observation can be used to describe the optimal decomposition of $(\alpha,\beta)$ into the points $S_i$. 
Sources in $\s_+$, starting from the source with the highest $p_i$, must contribute to the distance of $(\alpha,\beta)$ from the diagonal as much as possible. 
Therefore, starting from the source $\s_i$ with the highest $p_i$ and working downwards, we want to set as many $S_i = S_{H_i}$ as possible.
This situation splits into two more cases.
In the first one, we run out of sources in $\s_+$ before reaching the desired distance. This means that $\forall \s_i\in \s_+$, we have $S_i = S_{H_i}$, and the rest of the distance needs to be covered by sources in $S_-$. This can be done arbitrarily, as we have argued before.
This situation corresponds to the case $\sum_{i\in\s_+} \gamma_ip_i \leq \alpha-\beta$ in the proof in subsection 4.2 of the main text.
In the last case, the distance from the diagonal can be reached with $\s_i\in \s_+$ only. This leads to a situation where for the $k < |\s_+|$ sources with the highest $p_i$ we have that $S_i = S_{H_i}$, the point of the source with the $(k+1)$-st largest value of $p_i$ in $\s_+$ has a non-zero distance from the diagonal, and the rest of the points (of both $\s_+$ and $\s_-$) lie on the diagonal.This situation corresponds to the case $\sum_{i\in\s_+} \gamma_ip_i > \alpha-\beta$ in the proof in subsection 4.2 of the main text.
An example of the last case is depicted in the rightmost subfigure of Fig.~\ref{fig:mixedSourcePolytope}.
Note that while this argument helps to build an intuition, in order to fully solve the problem one must calculate the exact contributions of $\s_i\in\s_+$ to the distance of $(\alpha,\beta)$ from the diagonal (and thus the guessing entropy contributions) for an arbitrary feasible point of an arbitrary mixed source $\s$, in all three possible cases. This is formally done in subsection 4.2 of the main text.
}

\section{Mixed sources -- restrictions for partial solutions}

\label{appendix:PartialSolutionRestriction}

Here we discuss that if the observed statistics fulfill $\alpha\geq\beta$, then the optimal adversary's strategy fulfills $\alpha_{i}\geq\beta_{i}$, for all partial solutions $S_i$ associated with $\s_i$. This fact in turn implies that in the formulation of the optimization problem we can disregard strategies $\lambda_{i,{\lnot H}}$.

We prove the above by contradiction: assume that $\alpha \geq \beta$, but in the optimal solution some of the boxes $\s_{\uparrow}\subset \s$ have $\forall \s_j\in \s_{\uparrow}, S_j = (\alpha_j,\beta_j), \beta_j > \alpha_j$.
Let us define $\s_\downarrow = \s\backslash\s_\uparrow$. Then denote $S_{\uparrow} = (\alpha_{\uparrow}, \beta_{\uparrow})  = \frac{1}{\sum_{i\in \s_{\uparrow}}\gamma_i}\sum_{j \in \s_{\uparrow}} \gamma_j S_j$ and $S_\downarrow = (\alpha_\downarrow,\beta_\downarrow) = \frac{1}{\sum_{i \in \s\backslash\s_{\downarrow}} \gamma_i}\sum_{j \in \s_\downarrow} \gamma_j S_j$. 
The observed statistics can be now written as  $S = (\alpha,\beta) =   \sum_{i\in \s_{\uparrow}}\gamma_iS_{\uparrow} + \sum_{i \in \s_\downarrow} \gamma_i S_\downarrow$. 
Geometrically (see section \ref{appendix:Geometric interpretation}) this means that the point {$S$} lies between $S_{\uparrow}$ and $S_\downarrow$. 
This allows us to construct a new solution, in which both $S_{\uparrow}$ and $S_\downarrow$ are moved proportionally to $\gamma$ in the direction towards {$S$}. 
This decreases both $\alpha_\downarrow-\beta_\downarrow$ and $\beta_{\uparrow} - \alpha_{\uparrow}$.
Since the guessing probability of both partial solutions $S_\downarrow$ and $S_{\uparrow}$ directly depends on the distance from the diagonal, the new decomposition of $S_e$ has a lower guessing probability, which contradicts with the optimality of the original solution.


\section{Multiphoton events}

\label{appendix:MultiphotonEventsAreSimpleSources}
In this section we show that $n$ photon events in the setting described in
subsection 2.3 of the main text can be interpreted as simple sources with
$p=1-\pi^{n}$. Recall that in case the photon source emits $n$ photons, the
number of transmitted photons can vary between $0$ and $n$. The response
function of a measurement device needs to assign a click/no-click event to each received
photon number. This can be done in $2^{n+1}$ different ways.

Let us start by characterizing all the possible $2^{n+1}$ response functions.
Each response function can be characterized by specifying for which number of
transmitted photons the measurement device clicks, i.e.~by a subset $C\subseteq\{0,\dots,n\}$.
If we denote by $T$ the classical event of counting the transmitted
photons, we have that
\begin{align}
P(T=i) = {\binom{n}{i}}(1-\pi)^{i}\pi^{n-i}.
\end{align}
Now the response function indexed by $C$ can be characterized by the vector%

\begin{align}
S_{C} = \left(  P(\text{click}|x=0),P(\text{click}|x=1)\right)  = (p_{C},c),
\end{align}
where $p_{C} = \sum_{i\in C} P(T=i)$ and
\begin{align}
c=
\begin{cases}
1 \qquad\text{ if $0\in C$}\\
0\qquad\text{ if $0\notin C$.}%
\end{cases}
\end{align}
For each of these response functions, the adversary tries to guess whether the
measurement device clicks or does not click in a given round. The guessing probability corresponding to the response function $C$ is
\begin{align}
g_{C} = \max(p_{C},1-p_{C}). \label{guess,n,k}%
\end{align}

In what follows, we show that every strategy can be written as a convex
combination of the four response functions corresponding to $C=\emptyset$ (``Never
Click''), $C=\{0,\dots,n\}$, (``Always Click''), $C = \{1,\dots,n\}$ (``Honest
Strategy'') and $C = \{0\}$ (``Opposite of Honest Strategy''). Their
corresponding $S_{C}$ vectors are:
\begin{align}
S_{N}  &  = (0,0)\\
S_{Y}  &  = (1,1)\nonumber\\
S_{H}  &  = (1-\pi^{n},0)\nonumber\\
S_{\lnot H}  &  = (\pi^{n},1).\nonumber
\end{align}
Clearly, each response function with $0\notin C$ has $p_{C} \leq1-\pi^{n}$ and
therefore it can be expressed as a convex combination of $S_N$ and $S_{H}$.
Similarly, each response function with $0\in C$ has $p_{C} \geq\pi^{n}$ and
thus can be expressed as a convex combination of $S_Y$ and $S_{\lnot H}$.

In order to finish the argument, it remains to show that simulating each
$S_{C}$ with the above-mentioned strategies actually increases the adversary's
guessing probability. Without loss of generality let us examine response
functions with $0\notin C$. The argument for response functions with $0\in C$
is analogous. As argued before, any such strategy can be written as
\begin{align}
S_{C} = \lambda S_{H} + (1- \lambda)S_{N},
\end{align}
where $\lambda= \frac{p_{C}}{1-\pi^{n}}$. Recall that the guessing probability corresponding to $S_C$ is $g_{C} = \max
(p_{C},1-p_{C}) $. On the other hand, if $S_C$ is expressed as a convex combination of
$S_{H}$ and $S_{N}$, we have
\begin{align}
g^{\prime}_{C} = \lambda\left[  \max(\pi^{n},1-\pi^{n})\right]  + (1-\lambda).
\end{align}

For any strategy such that $0 \notin C$ we have $0\leq p_{C}\leq1-\pi^{n}$ and thus $\pi^{n}
\leq1-p_{C}\leq1$. If $1-\pi^{n} \leq\pi^{n}$, it also holds that
$p_{C}\leq1-p_{C}$ and
\begin{align}
g^{\prime}_{C}  &  = \lambda\left[  \max(\pi^{n},1-\pi^{n})\right]  +
(1-\lambda),\\
&  = \lambda\pi^{n} + (1-\lambda),\nonumber\\
&  = 1-\lambda(1-\pi^{n}),\nonumber\\
&  = 1- p_{C}\nonumber\\
&  = g_{C}.
\end{align}
In case $1-\pi^{n} \geq\pi^{n}$ we have
\begin{align}
g^{\prime}_{C}  &  = \lambda\left[  \max(\pi^{n},1-\pi^{n})\right]  +
(1-\lambda),\\
&  = \lambda(1-\pi^{n}) + (1-\lambda),\nonumber\\
&  = 1-\lambda\pi^{n}\nonumber\\
&  \geq g_{C},\nonumber
\end{align}
where the last inequality holds because $1-\lambda\pi^{n}$ is larger or equal
to both $1-p_{C} = 1-\lambda(1-\pi^{n})$ and $p_{C} = \lambda(1-\pi^{n})$.
This proves that expressing any strategy of the form $S_{C}$ as a convex
combination of $S_{H}$ and $S_{N}$ is not only possible, but advantageous for
the adversary. Therefore, the optimal adversary strategy in each $n$ photon event involves
only response functions that decide on whether the measurement device received a signal
(a positive number of photons) or not. Thus each $n$ photon event can be
described by a simple source $\mathcal{S}_i$, which sends a signal with probability $1-\pi^{n}$.

\section{Sampling error}\label{appendix:SamplingError}
In this section we address the inevitable uncertainty associated with estimating parameters of probability distributions. Both $\alpha = P(\text{click}| x=0)$ and $\beta = P(\text{click}| x=1)$ describe the parameters of Bernoulli random variables. Therefore, the most conservative approach is to use one-sided ($1-\epsilon$) confidence intervals to bound these parameters. Over-estimating $\beta$ or under-estimating $\alpha$ would lead to overestimating the entropy of the data. 

To see how this is done, first let $X_1, X_2, \ldots , X_n$ be a sequence of $n$ observed i.i.d. trials, with $X_i \sim \text{Bernoulli}(\theta)$, and $\bar{X} = \frac{1}{n}\sum_{i=1}^n X_i$. The desired error can be obtained by using the Chernoff--Hoeffding inequality.

\begin{align}\label{eq:Hoeffding1}
P(\bar{X}-E[\bar{X}]\geq t)&\leq e^{-2nt^2}\\\label{eq:Hoeffding2}
P(-\bar{X}+E[\bar{X}]\geq t)&\leq e^{-2nt^2}.
\end{align}

Note that since $E[\bar{X}] = \theta$, these inequalities bound the probability that the estimated (observed) probability of success $(\bar{X})$ is larger (\ref{eq:Hoeffding1}) or smaller (\ref{eq:Hoeffding2}) than the true value $\theta$ by more than $t$. By setting $\epsilon = e^{-2nt^2}$ and solving for $t$, we obtain $t=\sqrt{\ln{\left(\frac{1}{\epsilon}\right)}/2n}.$ Equipped with this, we construct bounded estimators $\hat{\alpha}$, $\hat{\beta}$, such that:
\begin{equation}
P\left(\alpha \in (\hat{\alpha},1] \And \beta \in [0,\hat{\beta}) \right) \geq (1-\epsilon)^2 \approx 1-2\epsilon.
\end{equation}
Since the choice of shutter settings are independent, and the trials themselves are i.i.d., the confidence of both estimators being in their respective intervals is the product of the individual events. During the experiment we observe {$t_{\alpha}$~($t_{\beta}$)} clicks in {$n_{\alpha}$ ($n_{\beta}$)} test rounds, from which we construct the estimators
\begin{align}
\hat{\alpha} &= \frac{t_\alpha}{n_\alpha} - \sqrt{\ln{\left(\frac{1}{\epsilon}\right)}/2n_\alpha} \\
\hat{\beta} &= \frac{t_\beta}{n_\beta} + \sqrt{\ln{\left(\frac{1}{\epsilon}\right)}/2n_\beta}.
\end{align}
The number of test rounds $n_\alpha$ and $n_\beta$ can be optimized as to increase the output entropy per batch of size $N$, by solving the following problem:
\begin{equation}
\label{eq:otpimtestrounds}
\max_{n_\alpha, n_\beta} \left[ H_{\text{min}}^*(\hat{\alpha},\hat{\beta})(N-n_\alpha-n_\beta) \right].
\end{equation}
Where $H_{\text{min}}^*$ is the calculated min-entropy per bit, depending on the assumed scenario. In practice, however, such an optimization requires the knowledge of the ratios $t_\alpha/n_\alpha$ and $t_\beta/n_\beta$, which are precisely the values that are being estimated. To break this cycle, this optimization can be done iteratively in practice, setting some original guesses (say $n=n_{\alpha,0}=n_{\beta,0}$) and on each subsequent batch of size $N$ assume that the clicking probabilities are the same as the previous round, which allows to solve equation \eqref{eq:otpimtestrounds}. It is important to highlight, however, that this optimization process is only for increasing the amount of extractable entropy, and the security of the protocol is not dependent on finding an optimal solution. Indeed, in our proof of principle experiment we only solved \eqref{eq:otpimtestrounds} approximately.

\end{document}